\documentstyle[12pt]{article}
\newcommand{\Cslash}{\not \!\! C}
\newcommand{\Dslash}{\not \!\! D}
\newcommand{\Aslash}{\not \!\! A}

\newcommand{\sslash}{\not \!\! s}
\newcommand{\pslash}{\not \!\! p}
\newcommand{\delslash}{\not \! \partial}
\renewcommand{\theequation}{\thesection.\arabic{equation}}

\newcommand{\sla}[1]{{\ooalign{\hfil/\hfil\crcr$#1$}}}
\newcommand{\f}[2]{\frac{#1}{#2}}

\newcommand{\lb}[0]{\l[}  % abbrev.of left bracket
\newcommand{\rb}[0]{\r]}
\newcommand{\lc}[0]{\l\{} % abbrev.of left curly bracket
\newcommand{\rc}[0]{\r\}}

\newcommand{\be}{\begin{eqnarray}}
\newcommand{\ee}{\end{eqnarray}}
\newcommand{\bc}{\begin{center}}
\newcommand{\ec}{\end{center}}

\def\l{\left}
\def\r{\right}

\def\hatmu{{\hat\mu}}

\def\wpk{\omega(p-k_1)}
\def\wp{\omega(p)}
\def\wq{\omega(q)}

\def\intp{\int\f{d^4p}{(2\pi)^4}}
\def\intpp{\int_{-\pi/a}^{\pi/a}\f{d^4p}{(2\pi)^4}}
\def\intq{\int\f{d^4q}{(2\pi)^4}}
\def\intt{\int\f{d^4t}{(2\pi)^4}}
\def\inttt{\int_{-\f{\pi}{a}}^{\f{\pi}{a}}\f{d^4t}{(2\pi)^4}}
\def\intk{\int_{-\f{\pi}{a}}^{\f{\pi}{a}}\f{d^4k}{(2\pi)^4}}
\def\intpqt{\int_{p,q,t}}
\def\ga{\gamma}
\def\g5{\gamma_5}

\begin{document}

\begin{flushright}{UT-987\\ hep-lat/0201016}
\end{flushright}
\vskip 0.5 truecm

\begin{center}
{\large{\bf A Perturbative Study of a General Class of Lattice 
Dirac Operators }}
\end{center}
\vskip .5 truecm
\centerline{\bf Kazuo Fujikawa and Masato Ishibashi}
\vskip .4 truecm
\centerline {\it Department of Physics,University of Tokyo}
\centerline {\it Bunkyo-ku,Tokyo 113,Japan}
\vskip 0.5 truecm

\makeatletter
\@addtoreset{equation}{section}
\def\theequation{\thesection.\arabic{equation}}
\makeatother

%\large
\begin{abstract}
A perturbative study of 
a general class of lattice Dirac operators is reported, which
 is based on an algebraic realization of the 
Ginsparg-Wilson relation in the form
 $\gamma_{5}(\gamma_{5}D)+(\gamma_{5}D)\gamma_{5} =
 2a^{2k+1}(\gamma_{5}D)^{2k+2}$ 
where $k$ stands for a non-negative integer. The choice  
$k=0$ corresponds to the commonly discussed Ginsparg-Wilson 
relation and thus to the overlap operator. 
We study one-loop fermion contributions to the 
self-energy of the gauge field, which are related to the 
fermion contributions to the one-loop $\beta$ function and to 
the Weyl anomaly. We first explicitly demonstrate that the Ward 
identity is satisfied by the self-energy tensor. By performing 
careful analyses, we then obtain the correct  
self-energy tensor free of infra-red divergences, as a general 
consideration of the Weyl anomaly indicates. This demonstrates 
that our general operators give correct chiral and Weyl 
anomalies. In general, however,  the Wilsonian 
effective action, which is supposed to be free of infra-red 
complications, is expected to be essential in the analyses of 
our general class of Dirac operators for dynamical gauge field. 
\end{abstract}

\section{Introduction}

Recent developments in the treatment of  fermions in lattice 
gauge theory are based on a  hermitian lattice Dirac  operator 
$\gamma_{5}D$ which satisfies the Ginsparg-Wilson relation[1]
\begin{equation}
\gamma_{5}D + D\gamma_{5} = 2aD\gamma_{5}D
\end{equation}
where the lattice spacing $a$ is utilized to make a dimensional
consideration transparent, and 
$\gamma_{5}$ is a hermitian chiral Dirac matrix. 
An explicit example of the operator satisfying (1.1) and free of 
species doubling has been given by Neuberger[2].  The relation 
(1.1) led to an interesting analysis of the notion of index in 
lattice gauge theory[3]. This index theorem in turn led to a 
new form of chiral symmetry, and the chiral anomaly is obtained 
as a non-trivial Jacobian factor under this modified chiral
 transformation[4]. This chiral Jacobian is regarded as 
a lattice realization of that in the continuum path integral[5]. 
See Refs.[6] for reviews of these developments. 

We have recently studied a specific generalization of the 
algebra (1.1)[7] 
\begin{equation}
\gamma_{5}(\gamma_{5}D)+(\gamma_{5}D)\gamma_{5}=
2a^{2k+1}(\gamma_{5}D)^{2k+2}
\end{equation}
where $k$ stands for a non-negative integer and $k=0$ corresponds
to the ordinary Ginsparg-Wilson relation. 
 When one defines 
\begin{equation}
H\equiv \gamma_{5}aD
\end{equation}
(1.2) is rewritten as 
\begin{equation}
\gamma_{5}H+H\gamma_{5}=2H^{2k+2}.
\end{equation}

The algebra (1.4) is equivalent to a set of equations
\begin{eqnarray}
&&H^{2k+1}\gamma_{5}+\gamma_{5}H^{2k+1}=2H^{2(2k+1)},\nonumber\\
&&H^{2}\gamma_{5}-\gamma_{5}H^{2}=0.
\end{eqnarray}
where the second relation is shown by
using the defining relation (1.4), and the first 
relation in  (1.5) becomes
identical to the ordinary Ginsparg-Wilson relation (1.1) if one 
defines
$H_{(2k+1)}=H^{2k+1}$. 
One can thus construct a solution to (1.5) by following the 
prescription used by Neuberger[2] 
\begin{equation}
\label{H_{(2k+1)}}
H_{(2k+1)} = \frac{1}{2}\gamma_{5}[ 1 + D_{W}^{(2k+1)}
\frac{1}{\sqrt{(D_{W}^{(2k+1)})^{\dagger}D_{W}^{(2k+1)}}}]
\end{equation} 
where 
\begin{equation}
D_{W}^{(2k+1)}\equiv i(\Cslash)^{2k+1}+B^{2k+1}
-(\frac{m_{0}}{a})^{2k+1}.
\end{equation}
Here we note that the conventional Wilson fermion operator 
$D_{W}$ ( with a non-zero mass term ) is given by
\begin{eqnarray}
D_{W}(x,y)&\equiv&i\gamma^{\mu}C_{\mu}(x,y)+B(x,y)-
\frac{1}{a}m_{0}\delta_{x,y},\nonumber\\
C_{\mu}(x,y)&=&\frac{1}{2a}[\delta_{x+\hat{\mu} a,y}
U_{\mu}
(y)-\delta_{x,y+\hat{\mu} a}U^{\dagger}_{\mu}(x)],
\nonumber\\
B(x,y)&=&\frac{r}{2a}\sum_{\mu}[2\delta_{x,y}-
\delta_{y+\hat{\mu} a,x}U_{\mu}^{\dagger}(x)
-\delta_{y,x+\hat{\mu} a}U_{\mu}(y)],
\nonumber\\
U_{\mu}(y)&=& \exp [iagA_{\mu}(y)].
\end{eqnarray}
 The parameter $r$ stands for the Wilson parameter.
Our matrix convention is that $\gamma^{\mu}$ are anti-hermitian, 
$(\gamma^{\mu})^{\dagger} = - \gamma^{\mu}$, and thus 
$\Cslash\equiv \gamma^{\mu}C_{\mu}(n,m)$ is hermitian
\begin{equation}
\Cslash^{\dagger} = \Cslash.
\end{equation} 

The hermitian operator $H$ itself is then finally defined by 
(in the representation where $H_{(2k+1)}$ is diagonal)
\begin{equation}
H=(H_{(2k+1)})^{1/2k+1}
\end{equation}
in such a manner that the second relation of (1.5) is satisfied,
 which is in fact confirmed in the representation where 
$H_{(2k+1)}$ is diagonal[7].  Also the conditions $0<m_{0}<2r=2$
and 
\begin{equation}
2m_{0}^{2k+1}=1
\end{equation}
ensure the absence of species doublers and a proper 
normalization of the Dirac operator $H$.

The locality properties are crucial in any construction of 
lattice Dirac operator, and the  locality of the standard overlap
operator with $k=0$ has been established by Hernandez, Jansen 
and L\"{u}scher[10], and by Neuberger[11].   

As for the direct proof of locality of the operator $D$ for 
general $k$, it is shown for the 
vanishing gauge field by using the explicit solution
for the operator $H$ in momentum representation[12][9]
\begin{eqnarray}
H(ap_{\mu})&=&\gamma_{5}(\frac{1}{2})^{\frac{k+1}{2k+1}}
(\frac{1}{\sqrt{H^{2}_{W}}})^{\frac{k+1}{2k+1}}
\{(\sqrt{H^{2}_{W}}+M_{k})^{\frac{k+1}{2k+1}}
-(\sqrt{H^{2}_{W}}-M_{k})^{\frac{k}{2k+1}}
\frac{\sslash}{a} \}\nonumber\\
&=&\gamma_{5}(\frac{1}{2})^{\frac{k+1}{2k+1}}
(\frac{1}{\sqrt{F_{(k)}}})^{\frac{k+1}{2k+1}}
\{(\sqrt{F_{(k)}}+\tilde{M}_{k})^{\frac{k+1}{2k+1}}
-(\sqrt{F_{(k)}}-\tilde{M}_{k})^{\frac{k}{2k+1}}
\sslash \}\nonumber\\
&&
\end{eqnarray}
where
\begin{eqnarray}
F_{(k)}&=&(s^{2})^{2k+1}+\tilde{M}_{k}^{2},\nonumber\\
\tilde{M}_{k}&=&[\sum_{\mu}(1-c_{\mu})]^{2k+1}
-m_{0}^{2k+1}
\end{eqnarray}
and
\begin{eqnarray}
&&s_{\mu}=\sin ap_{\mu}\nonumber\\
&&c_{\mu}=\cos ap_{\mu}\nonumber\\
&&\sslash=\gamma^{\mu}\sin ap_{\mu}.
\end{eqnarray}
For $k=0$, this operator is reduced to  Neuberger's overlap 
operator[2]. 
Here the inner product is defined to be $s^{2}\geq 0$.
This operator for an infinitesimal $p_{\mu}$, i.e.,
for $|ap_{\mu}|\ll 1$, gives rise to 
\begin{equation}
H\simeq-\gamma_{5}a\pslash(1+O(ap)^{2})
+\gamma_{5}(\gamma_{5}a\pslash)^{2k+2}
\end{equation}
to be consistent with $H=\gamma_{5}aD$; the last term in the 
right-hand side is the 
leading term of chiral symmetry breaking terms.
The locality of this explicit construction (1.12) has been shown 
by examining the analytic properties in the Brillouin zone[12]. 

It is 
important to recognize that this operator is not ultra-local but 
exponentially local[13]; the operator $H(x,y)$ in (1.12) decays 
exponentially 
for large separation in coordinate representation as[12]
\begin{equation}
H(x,y)\sim \exp[-|x-y|/(2.5ka)].
\end{equation}
An explicit analysis of the locality of the operator 
$H_{(2k+1)}=H^{2k+1}$ ({\em not} $H$ itself ) in the presence of 
gauge field, in particular, the locality 
domain for the gauge field strength $||F_{\mu\nu}||$ has been 
performed. The locality domain for $||F_{\mu\nu}||$ becomes 
smaller for larger $k$, 
but a definite non-zero domain has been established[12]. The 
remaing task is to show the locality of  the 
operator $H=(H_{(2k+1)})^{1/(2k+1)}$ itself in the presence of
gauge field. Due to the operation of taking the $(2k+1)$th root,
 an explicit analysis has not been performed yet, though a 
supporting argument has been given in Ref.[12].

It has been shown that all the good chiral properties of 
the overlap operator[2] are retained in the generalization in
(1.4)[8][9]. The practical applications of this generalized 
operator $D$ are not known at this moment.
We however mention the characteristic properties of this 
generalization: The spectrum near the continuum configuration 
is closer to that of continuum theory and the chiral symmetry 
breaking terms become more irrelevant in the continuum limit
for $k>0$. The operator however spreads over more lattice
points for larger $k$, as is indicated in (1.16).

In this paper we study a perturbative aspect of the general 
class of Dirac operators. To be specific, we study the 
one-loop fermion contribution to the gauge field self-energy, 
which is related to the $\beta$ function and to the Weyl anomaly.

\section{Self-energy tensor, $\beta$-function and Weyl anomaly}

The lattice perturbation theory is very tedious in 
general[14]-[22], and
it is more so in our generalization. For this reason, we study
the simplest diagrams related to the one-loop self-energy 
correction to gauge fields. This effect is also related to the 
fermion contribution to the lowest order $\beta$-function and 
to the Weyl anomaly [23][24].  
A rather general analysis of Weyl anomaly is possible, 
and we first briefly summarize it.

In the standard continuum formulation, one starts with the 
path integral defined in a background curved space[25]
\begin{equation}
\int d\mu\exp[\int d^{4}x\sqrt{g}\bar{\psi}i\Dslash\psi].
\end{equation}
The general coordinate invariant path integral measure is 
defined by 
\begin{equation}
d\mu={\cal D}\tilde{\bar{\psi}}{\cal D}\tilde{\psi}
\end{equation}
and the Weyl transformation laws are given by
\begin{eqnarray}
&&e_{a}^{\mu}(x)\rightarrow \exp[\alpha(x)]e_{a}^{\mu}(x),
\nonumber\\
&&\tilde{\psi}(x)\equiv (g)^{1/4}\psi(x)\rightarrow 
\exp[-\frac{1}{2}\alpha(x)]
\tilde{\psi}(x),\nonumber\\
&&\tilde{\bar{\psi}}(x)\equiv (g)^{1/4}\bar{\psi}(x)\rightarrow 
\exp[-\frac{1}{2}\alpha(x)]\tilde{\bar{\psi}}(x)
\end{eqnarray}
where $e_{a}^{\mu}(x)$ stands for the vierbein. This 
transformation law is fixed by the invariance of the action in
the above path integral
under a global (i.e., constant) $\alpha$, and the Weyl weight 
factor of fermionic variables is essentially defined by the 
vierbein in $\Dslash=e_{a}^{\mu}(x)\gamma^{a}D_{\mu}$.

The Jacobian for this transformation of fermionic variables is 
given by
\begin{equation}
\ln J(\alpha)=\lim_{M\rightarrow\infty}
Tr\alpha(x)\exp[-(\Dslash/M)^{2}]=Tr\alpha(x)
\frac{g^{2}}{24\pi^{2}}F_{\mu\nu}F^{\mu\nu}
\end{equation}
where the mode cut-off of $\Dslash$ is provided by 
$e^{-(\lambda_{n}/M)^{2}}$ in terms of the eigenvalues of 
$\Dslash$. See Ref.[25] for further details.

When one analyzes the higher derivative theory
\begin{equation}
{\cal L}_{2k+1}=\int d^{4}x\sqrt{g}\bar{\psi}i(\Dslash)^{2k+1}
\psi,
\end{equation}
the Weyl transformation laws are given by
\begin{eqnarray}
&&e_{a}^{\mu}(x)\rightarrow \exp[\alpha(x)]e_{a}^{\mu}(x),
\nonumber\\
&&\tilde{\psi}(x)=(g)^{1/4}\psi(x)\rightarrow 
\exp[-\frac{2k+1}{2}\alpha(x)]
\tilde{\psi}(x),\nonumber\\
&&\tilde{\bar{\psi}}(x)=(g)^{1/4}\bar{\psi}(x)\rightarrow 
\exp[-\frac{2k+1}{2}\alpha(x)]\tilde{\bar{\psi}}(x)
\end{eqnarray}
and the Weyl anomaly is given by
\begin{equation}
\ln J_{2k+1}(\alpha)=\lim_{M\rightarrow\infty}
Tr(2k+1)\alpha(x)\exp[-((\Dslash)^{2k+1}/M^{2k+1})^{2}].
\end{equation}
Since the Weyl anomaly is independent of the regulator 
function [25],
we have
\begin{equation}
\ln J_{2k+1}(\alpha)=(2k+1)\ln J(\alpha).
\end{equation}

This relation (2.8) is also understood from a view point of the 
self-energy correction to the gauge field as follows:
\begin{eqnarray}
\det (\Dslash)^{2k+1}&=&\exp[(2k+1)Tr\ln\Dslash]\nonumber\\
&=&\exp[(2k+1)Tr\ln(\delslash-ig\Aslash)]\\
&=&\exp[(2k+1)Tr\ln\delslash-\frac{2k+1}{2}(ig)^{2}
Tr\frac{1}{\delslash}\Aslash\frac{1}{\delslash}\Aslash +....]
\nonumber
\end{eqnarray}
The term quadratic in the gauge field $A_{\mu}$ gives the 
self-energy correction, which is $2k+1$ times larger than the 
self-energy correction generated by $\det \Dslash$.

This analysis of the self-energy correction is applicable to 
the present lattice operator. By our definition in (1.10)
we have 
\begin{equation}
\exp[Tr\ln H]=\exp[Tr\ln H_{(2k+1)}^{1/(2k+1)}]
=\exp[\frac{1}{2k+1}Tr \ln H_{(2k+1)}].
\end{equation}
For a sufficiently small coupling constant $g$, we have 
\begin{equation}
\exp[Tr \ln H]=\exp[Tr \ln H^{(0)}
+Trg^{2}A_{\mu}(x)O(x,y)_{\mu\nu}A_{\nu}(y)+O(g^{3})]
\end{equation}
where $H^{(0)}$ stands for the free Dirac operator given in 
(1.12), and the second term stands for the lowest order term in 
the effective potential and thus for the lowest order 
self-energy correction to the gauge field. Similarly, we have   
\begin{eqnarray}
&&\exp[\frac{1}{2k+1}Tr\ln H_{(2k+1)}]
\\
&=&\exp[\frac{1}{2k+1}Tr\ln H^{(0)}_{(2k+1)}+ \frac{1}{2k+1}
Trg^{2}A_{\mu}(x)\tilde{O}(x,y)_{\mu\nu}A_{\nu}(y)+O(g^{3})]. 
\nonumber
\end{eqnarray}
where $Tr\ln H^{(0)}_{(2k+1)}$ stands for the free part of 
$H_{(2k+1)}$. Those zeroth order terms satisfy the relation 
\begin{equation}
Tr \ln H^{(0)}=\frac{1}{2k+1}Tr\ln H^{(0)}_{(2k+1)}
\end{equation}
if one uses the explicit form of the operator in (1.12).

We thus conclude
\begin{equation}
Trg^{2}A_{\mu}(x)O(x,y)_{\mu\nu}A_{\nu}(y)
=\frac{1}{2k+1}Trg^{2}A_{\mu}(x)\tilde{O}(x,y)_{\mu\nu}A_{\nu}(y)
\end{equation}
for a sufficiently small coupling constant $g$, which shows that
the lowest order self-energy correction in the left-hand side 
for the operator $H$ is evaluated by the self-energy correction
in terms of $H_{(2k+1)}$. We use this relation for the 
evaluation of the lowest order self-energy correction for any 
$k\geq 1$. Note that the operator $H_{(2k+1)}=H^{2k+1}$ is much 
better understood than $H$ itself in our construction.
We also confirm that this relation (2.14) is in fact valid by 
evaluating the left-hand side directly for the simplest case 
$k=1$.    

>From a view point of Weyl anomaly, one may tentatively take 
\begin{eqnarray}
&&\lim_{M\rightarrow\infty}
Tr(2k+1)\alpha(x)\exp[-(H_{(2k+1)}/(aM)^{2k+1})^{2}]\nonumber\\
&&\rightarrow
\lim_{M\rightarrow\infty}
Tr(2k+1)\alpha(x)\exp[-((\Dslash)^{2k+1}/M^{2k+1})^{2}]   
\end{eqnarray}
as a lattice version of the Weyl anomaly. We then obtain the 
same result as the self-energy correction in the limit 
$a\rightarrow 0$, although no systematic formulation of Weyl 
anomaly on the lattice is known.

In lattice perturbative calculations, however, we should be 
careful of the possible appearance of infrared divergences, 
which should cancel in the final result. We show that  a 
careful analysis gives the correct result of continuum theory 
free of infrared divergences for $a\rightarrow 0$. 

\section{The vacuum polarization tensor by $H_{(2k+1)}$}

In this section, we calculate the one-loop fermion contribution
to the vacuum polarization  $\Pi_{\mu\nu}$ on the basis of the 
operator $H_{(2k+1)}$ ( and not $H$ itself) following the 
analyses in Section 2. We first  show that
the Ward identity is satisfied to be consistent with gauge 
invariance and that there appear no  divergences except for 
the logarithmic divergence for $a\to 0$. We then discuss the 
gauge field wave function renormalization factor.

Feynman diagrams for the vacuum polarization with  
fermion one-loop are shown in Fig.1, and the necessary Feynman 
rules are given in Appendix A.

\begin{figure}[htbp]
\begin{center}
%WinTpicVersion2.15
\unitlength 0.1in
\begin{picture}(50.10,18.60)(2.00,-24.30)
% CIRCLE 2 0 3 0
% 4 1200 1820 1480 2170 1480 2170 1480 2170
% 
\special{pn 8}%
\special{ar 1200 1420 448 448  0.0000000 6.2831853}%
% LINE 2 2 3 0
% 2 200 1800 750 1800
% 
\special{pn 8}%
\special{pa 200 1400}%
\special{pa 750 1400}%
\special{dt 0.045}%
\special{pa 750 1400}%
\special{pa 749 1400}%
\special{dt 0.045}%
% LINE 2 2 3 0
% 2 1650 1800 2210 1800
% 
\special{pn 8}%
\special{pa 1650 1400}%
\special{pa 2210 1400}%
\special{dt 0.045}%
\special{pa 2210 1400}%
\special{pa 2209 1400}%
\special{dt 0.045}%
% LINE 2 2 3 0
% 2 3210 2270 5210 2270
% 
\special{pn 8}%
\special{pa 3210 1870}%
\special{pa 5210 1870}%
\special{dt 0.045}%
\special{pa 5210 1870}%
\special{pa 5209 1870}%
\special{dt 0.045}%
% CIRCLE 2 0 3 0
% 4 4200 1800 4200 2270 4200 2270 4200 2270
% 
\special{pn 8}%
\special{ar 4200 1400 470 470  0.0000000 6.2831853}%
% VECTOR 2 0 3 0
% 2 1150 1290 1330 1290
% 
\special{pn 8}%
\special{pa 1150 890}%
\special{pa 1330 890}%
\special{fp}%
\special{sh 1}%
\special{pa 1330 890}%
\special{pa 1263 870}%
\special{pa 1277 890}%
\special{pa 1263 910}%
\special{pa 1330 890}%
\special{fp}%
% VECTOR 2 0 3 0
% 2 1280 2340 1070 2340
% 
\special{pn 8}%
\special{pa 1280 1940}%
\special{pa 1070 1940}%
\special{fp}%
\special{sh 1}%
\special{pa 1070 1940}%
\special{pa 1137 1960}%
\special{pa 1123 1940}%
\special{pa 1137 1920}%
\special{pa 1070 1940}%
\special{fp}%
% VECTOR 2 0 3 0
% 2 390 1700 590 1700
% 
\special{pn 8}%
\special{pa 390 1300}%
\special{pa 590 1300}%
\special{fp}%
\special{sh 1}%
\special{pa 590 1300}%
\special{pa 523 1280}%
\special{pa 537 1300}%
\special{pa 523 1320}%
\special{pa 590 1300}%
\special{fp}%
% VECTOR 2 0 3 0
% 2 1810 1700 2030 1700
% 
\special{pn 8}%
\special{pa 1810 1300}%
\special{pa 2030 1300}%
\special{fp}%
\special{sh 1}%
\special{pa 2030 1300}%
\special{pa 1963 1280}%
\special{pa 1977 1300}%
\special{pa 1963 1320}%
\special{pa 2030 1300}%
\special{fp}%
% VECTOR 2 0 3 0
% 2 3420 2150 3590 2150
% 
\special{pn 8}%
\special{pa 3420 1750}%
\special{pa 3590 1750}%
\special{fp}%
\special{sh 1}%
\special{pa 3590 1750}%
\special{pa 3523 1730}%
\special{pa 3537 1750}%
\special{pa 3523 1770}%
\special{pa 3590 1750}%
\special{fp}%
% VECTOR 2 0 3 0
% 2 4810 2160 5000 2160
% 
\special{pn 8}%
\special{pa 4810 1760}%
\special{pa 5000 1760}%
\special{fp}%
\special{sh 1}%
\special{pa 5000 1760}%
\special{pa 4933 1740}%
\special{pa 4947 1760}%
\special{pa 4933 1780}%
\special{pa 5000 1760}%
\special{fp}%
% VECTOR 2 0 3 0
% 2 4290 1250 4120 1250
% 
\special{pn 8}%
\special{pa 4290 850}%
\special{pa 4120 850}%
\special{fp}%
\special{sh 1}%
\special{pa 4120 850}%
\special{pa 4187 870}%
\special{pa 4173 850}%
\special{pa 4187 830}%
\special{pa 4120 850}%
\special{fp}%
% STR 2 0 3 0
% 3 200 1870 200 1970 1 0
% p,$\mu$,A
\put(2.0000,-15.7000){\makebox(0,0)[lt]{p,$\mu$,A}}%
% STR 2 0 3 0
% 3 1730 1880 1730 1980 1 0
% p,$\nu$,B
\put(17.3000,-15.8000){\makebox(0,0)[lt]{p,$\nu$,B}}%
% STR 2 0 3 0
% 3 1060 930 1060 1030 1 0
% t+p
\put(10.6000,-6.3000){\makebox(0,0)[lt]{t+p}}%
% STR 2 0 3 0
% 3 1140 2360 1140 2460 1 0
% t
\put(11.4000,-20.6000){\makebox(0,0)[lt]{t}}%
% STR 2 0 3 0
% 3 4130 870 4130 970 1 0
% t
\put(41.3000,-5.7000){\makebox(0,0)[lt]{t}}%
% STR 2 0 3 0
% 3 3190 2270 3190 2370 1 0
% p,$\mu$,A
\put(31.9000,-19.7000){\makebox(0,0)[lt]{p,$\mu$,A}}%
% STR 2 0 3 0
% 3 4660 2280 4660 2380 1 0
% p,$\nu$,B
\put(46.6000,-19.8000){\makebox(0,0)[lt]{p,$\nu$,B}}%
% STR 2 0 3 0
% 3 1010 2730 1010 2830 1 0
% (a)
\put(10.1000,-24.3000){\makebox(0,0)[lt]{(a)}}%
% STR 2 0 3 0
% 3 4130 2720 4130 2820 1 0
% (b)
\put(41.3000,-24.2000){\makebox(0,0)[lt]{(b)}}%
\end{picture}%
\end{center}
\label{fig3}
\caption{Feynman diagrams for the vacuum polarization}
\end{figure}
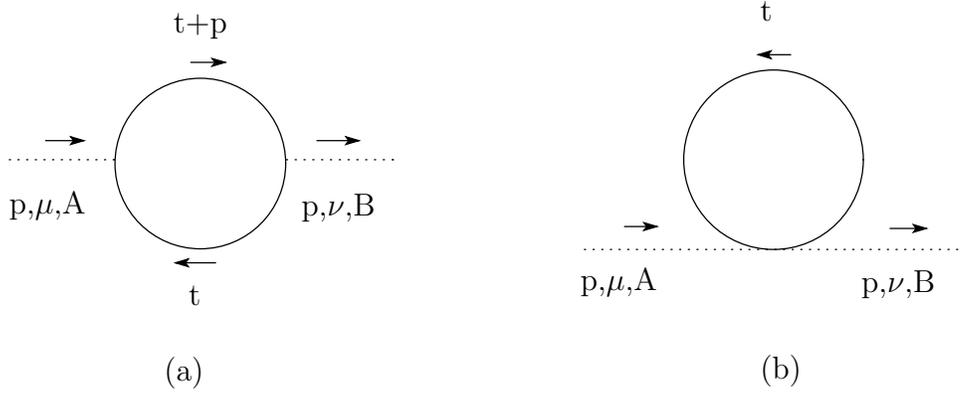

The amplitude corresponding to Fig.1(a) is given in terms of 
the notation in Appendix A by (by using $tr(T^A T^B)
=\frac{1}{2}\delta^{AB}$ and $N_{f}$ flavors in QCD)
\begin{eqnarray}
\label{pia1}
&&\Pi_{\mu\nu}^{(a)}(p)=\f{-g^2}{4a^{4k+2}}
\frac{N_{f}}{2}\delta^{AB}
\int_t
\f{1}{\lc w(t)+w(t+p)\rc^2}\nonumber\\
&&\times
tr\lb D_0^{-1}(t)\g5\lc X_{1\mu}(t,t+p,-p)-\f{X_0(t)}{w(t)}
X_{1\mu}^\dagger(t,t+p,-p)\f{X_0(t+p)}{w(t+p)}\rc\right.
\nonumber\\
&&\left.\times D_0^{-1}(t+p)\g5
\lc X_{1\nu}(t+p,t,p)-\f{X_0(t+p)}{w(t+p)}
X_{1\nu}^\dagger(t+p,t,p)\f{X_0(t)}{w(t)}\rc\rb.\nonumber\\ 
\end{eqnarray}
We omit the factor $\delta^{AB}$ from now on.

The amplitude corresponding to Fig.1(b) is similarly given by
\begin{eqnarray}
\label{pib1}
&&\Pi_{\mu\nu}^{(b)}(p)=\f{g^2}{2a^{2k+1}}\frac{N_{f}}{2}
\int_t
tr\lb D_0^{-1}(t)\g5\f{1}{2w(t)}\right.\nonumber\\
&&\times\lb X_{2\mu\nu}(t,t,-p,p) - \f{X_0(t)}{w(t)}
X^\dagger_{2\mu\nu}(t,t,-p,p)\f{X_0(t)}{w(t)}\right.\nonumber\\
&&-\f{1}{(w(t)+w(t+p))^2}\nonumber\\
&&\times\lc X_{1\mu}(t,p+t,-p)
X^\dagger_{1\nu}(p+t,t,p)X_0(t)\right.\nonumber\\
&&+X_{1\mu}(t,p+t,-p)
X_0^\dagger(p+t)X_{1\nu}(p+t,t,p)\nonumber\\
&&+X_0(t)X^\dagger_{1\mu}(t,p+t,-p)
X_{1\nu}(p+t,t,p)\nonumber\\
&&-\f{2w(t)+w(p+t)}{w(t)^2w(p+t)}\nonumber\\
&&\left.\times X_0(t)X^\dagger_{1\mu}(t,p+t,-p)
X_0(p+t)X^\dagger_{1\nu}(p+t,t,p)X_0(t)\rc\nonumber\\
&& +X_{2\nu\mu}(t,t,p,-p) - \f{X_0(t)}{w(t)}
X^\dagger_{2\nu\mu}(t,t,p,-p)\f{X_0(t)}{w(t)}\nonumber\\
&&-\f{1}{(w(t)+w(t-p))^2}\nonumber\\
&&\times\lc X_{1\nu}(t,t-p,p)
X^\dagger_{1\mu}(t-p,t,-p)X_0(t)\right.\nonumber\\
&&+X_{1\nu}(t,t-p,p)
X_0^\dagger(t-p)X_{1\mu}(t-p,t,-p)\nonumber\\
&&+X_0(t)X^\dagger_{1\nu}(t,t-p,p)
X_{1\mu}(t-p,t,-p)\nonumber\\
&&-\f{2w(t)+w(t-p)}{w(t)^2w(t-p)}\nonumber\\
&&\left.\left.\left.\times X_0(t)X^\dagger_{1\nu}(t,t-p,p)
X_0(t-p)X^\dagger_{1\mu}(t-p,t,-p)X_0(t)\rc\rb\rb.\nonumber\\
\end{eqnarray}  

\subsection{Ward identity}

We first show that
the Ward identity for $\Pi_{\mu\nu}$ as a manifestation of 
gauge invariance holds as follows,
\begin{eqnarray}
\label{Ward id}
\sum_{\nu}\tilde{p}_\nu\left(\Pi_{\mu\nu}^{(a)}(p)
+\Pi_{\mu\nu}^{(b)}(p))\right)&=&0,
\end{eqnarray}
where $\tilde{p}_\nu = (2/a)\sin ap_\nu/2$.\footnote{The Ward
identity in the case of the overlap Dirac operator has been
confirmed explicitly in Ref.\cite{kikukawanoguchi}.} 
For this purpose we first calculate 
$\sum_{\nu}\tilde{p}_\nu X_{1\nu}$ 
and $\sum_{\nu}\tilde{p}_\nu X_{2\mu\nu}$. 
For $\sum_{\nu}\tilde{p}_\nu X_{1\nu}$ we have
\begin{eqnarray}
&&a^{2k+1}\sum_{\nu}\tilde{p}_\nu X_{1\nu}(t+p,t,p)\nonumber\\
&&=\sum_{l+m=2k}\lb i(i\sla{s}_{t+p})^l
(i\sla{s}_{t+p}-i\sla{s}_{t})i(i\sla{s}_t)^m
\right.\nonumber\\
&&+\left(r\sum_{\rho}(1-\cos (t+p)_\rho a)\right)^l
\lc\left(r\sum_{\rho}(1-\cos (t+p)_\rho a)\right)
-\left(r\sum_{\rho}(1-\cos t_\rho a)\right)\rc\nonumber\\
&&\left.\times \left(r\sum_{\rho}(1-\cos t_\rho a)\right)^m\rb
\nonumber\\
&&=i(i\sla{s}_{t+p})^{2k+1} 
+ \left(r\sum_{\rho}(1-\cos (t+p)_\rho a)
\right)^{2k+1} - m_0^{2k+1}\nonumber\\
&&-\lc i(i\sla{s}_t)^{2k+1} + 
\left(r\sum_{\rho}(1-\cos t_\rho a)
\right)^{2k+1} - m_0^{2k+1}\rc\nonumber\\
&&=a^{2k+1}\lc X_0(t+p) - X_0(t) \rc,
\end{eqnarray}  
and further we have
\begin{eqnarray}
&&\sum_{\nu}\tilde{p}_\nu\left(X_{1\nu}(t+p,t,p)
-\f{X_0(t+p)}{w(t+p)}X_{1\nu}^\dagger(t+p,t,p)\f{X_0(t)}{w(t)}
\right)
\nonumber\\
&&=2\g5 a^{2k+1}(w(t+p)+w(t))(D_0(p+t)-D_0(t)),
\end{eqnarray}
where we used the following relations 
\begin{eqnarray}
&&\sum_{\nu}\tilde{p}_\nu \ga_\nu\cos (t+p/2)_\nu a
=\f{1}{a}(\sla{s}_{t+p} -\sla{s}_t),\\
&&\sum_{\nu}\tilde{p}_\nu r
\sin(t+p/2)_\nu a = \f{1}{a}\lc
\left(r\sum_{\rho}(1-\cos (t+p)_\rho a)
\right)-\left(r\sum_{\rho}(1-\cos t_\rho a)\right)\rc.
\nonumber\\ 
\end{eqnarray}
Using the above relations, we also have
\begin{eqnarray}
&&a^{2k}\sum_{\nu}\tilde{p}_\nu X_{2\mu\nu}(t,t,-p,p)\nonumber\\
&&=\sum_{l+m=2k}\lb 
i(i\sla{s}_t)^l\left(i\gamma_\mu \cos \left(t+\f{p}{2}\right)_\mu
a\right)(i\sla{s}_{t+p})^m\right.\nonumber\\
&&-i(i\sla{s}_t)^l\left(i\gamma_\mu \cos \left(t+\f{p}{2}
\right)_\mu
a\right)(i\sla{s}_{t})^m\nonumber\\
&&+\left(r\sum_{\rho}(1-\cos t_\rho a)
\right)^l\left(r\sin\left(t+\f{p}{2}\right)_\mu a\right)
\left(r\sum_{\rho}(1-\cos (t+p)_\rho a)
\right)^m\nonumber\\
&&-\left(r\sum_{\rho}(1-\cos t_\rho a)
\right)^l\left(r\sin\left(t+\f{p}{2}\right)_\mu a\right)
\left(r\sum_{\rho}(1-\cos t_\rho a)
\right)^m\nonumber\\
&&+\f{1}{2}\lc i
\left(i\sla{s}_t\right)^l\left(i\ga_\mu
\cos\left(t+\f{p}{2}\right)_\mu a -i\ga_\mu
\cos\left(t-\f{p}{2}\right)_\mu a\right)
\left(i\sla{s}_t\right)^{m}\right.\nonumber\\
&& +
\left(r\sum_\rho(1-\cos t_\rho a)\right)^{l}
\left(r\sin\left(t+\f{p}{2}\right)_\mu a
-r\sin\left(t-\f{p}{2}\right)_\mu a \right)\nonumber\\
&&\left.\left.\times\left(r\sum_\rho (1-\cos t_\rho a)
\right)^{m}\rc\rb.\nonumber\\
\end{eqnarray}
Therefore we obtain
\begin{eqnarray}
&&\sum_{\nu}\tilde{p}_\nu[X_{2\mu\nu}(t,t,-p,p)
+X_{2\nu\mu}(t,t,p,-p)]\nonumber\\
&&=X_{1\mu}(t,t+p,-p) -X_{1\mu}(t-p,t,-p).
\end{eqnarray} 
By using these relations, $\sum_{\nu}\tilde{p}_\nu 
\Pi_{\mu\nu}^{(a)}(p)$ is written as 
\begin{eqnarray}
&&\sum_{\nu}\tilde{p}_\nu 
\Pi_{\mu\nu}^{(a)}(p)=\f{-g^2}{2a^{2k+1}}\frac{N_{f}}{2}
\int_t
\f{1}{\lc w(t)+w(t+p)\rc}tr\lb \lc D_0^{-1}(t)-
D_0^{-1}(t+p)\rc\g5\right.\nonumber\\
&&\left.\times\lc X_{1\mu}(t,t+p,-p)-\f{X_0(t)}{w(t)}
X_{1\mu}^\dagger(t,t+p,-p)\f{X_0(t+p)}{w(t+p)}\rc\rb,
\end{eqnarray}
and similarly $\sum_{\nu}\tilde{p}_\nu 
\Pi_{\mu\nu}^{(b)}(p)$ is written as 
\begin{eqnarray}
&&\sum_{\nu}\tilde{p}_\nu 
\Pi_{\mu\nu}^{(b)}(p)=\f{g^2}{2a^{2k+1}}\frac{N_{f}}{2}
\int_t
tr\lb D_0^{-1}(t)\g5\right.\nonumber\\
&&\times\lb \f{1}{w(t)+w(t+p)}\lc X_{1\mu}(t,t+p,-p) 
- \f{X_0(t)}{w(t)}
X^\dagger_{1\mu}(t,t+p,-p)\f{X_0(t+p)}{w(t+p)}\rc\right.
\nonumber\\
&&\left.\left.-\f{1}{w(t)+w(t-p)}\lc X_{1\mu}(t-p,t,-p) 
- \f{X_0(t-p)}{w(t-p)}
X^\dagger_{1\mu}(t-p,t,-p)\f{X_0(t)}{w(t)}\rc\rb\rb\nonumber\\
&&=\f{g^2}{2a^{2k+1}}\f{N_f}{2}\int_t
tr\lb \lc D_0^{-1}(t)-D_0^{-1}(t+p)\rc\g5\right.\nonumber\\
&&\times\lb \f{1}{w(t)+w(t+p)}\lc X_{1\mu}(t,t+p,-p) 
- \f{X_0(t)}{w(t)}
X^\dagger_{1\mu}(t,t+p,-p)\f{X_0(t+p)}{w(t+p)}\rc\rb.
\nonumber\\
\end{eqnarray}
Combining these two expressions,  the Ward identity for the 
vacuum polarization tensor holds as in Eq.(\ref{Ward id}). This 
Ward identity
dictates  the tensor structure of $\Pi_{\mu\nu}(p)$ for 
small $p_\mu$ to be 
\begin{eqnarray}
\label{vptensor}
\Pi_{\mu\nu}(p)\simeq(p^2\delta_{\mu\nu}-p_\mu
p_\nu)\Pi(a^2p^2).
\end{eqnarray}

\subsection{Structure of  divergences}

We next examine the structure of various divergences. To 
evaluate the divergent parts of $\Pi_{\mu\nu}^{(a)}(p)$ and  
$\Pi_{\mu\nu}^{(b)}(p)$, we rescale the integration momenta
$t_\mu \to t_\mu/a$ in each amplitude (\ref{pia1}) and 
(\ref{pib1}). For QCD with $N_{f}$ flavors, we obtain 
\begin{eqnarray}
\label{pia}
&&\Pi_{\mu\nu}^{(a)}(p)=\f{-N_{f} g^2}{8a^2}\int_t
\f{1}{\lc w(t)+w(t+pa)\rc^2}\nonumber\\
&&tr\lb\lc \f{(s_t^2)^k\sla{s}_t}{w(t)+M(t)} + 1\rc\right.
\nonumber\\
&&\times \lc X_{1\mu}(t,t+pa,-pa)-\f{X_0(t)}{w(t)}
X_{1\mu}^\dagger(t,t+pa,-pa)\f{X_0(t+pa)}{w(t+pa)}\rc
\nonumber\\
&&\times \lc \f{(s_{t+pa}^2)^k\sla{s}_{t+pa}}{w(t+pa)+M(t+pa)} 
+ 1\rc
\nonumber\\
&&\left.\times \lc X_{1\nu}(t+pa,t,pa)-\f{X_0(t+pa)}{w(t+pa)}
X_{1\nu}^\dagger(t+pa,t,pa)\f{X_0(t)}{w(t)}\rc\rb.\nonumber\\ 
\end{eqnarray}
and
\begin{eqnarray}
\label{pib}
&&\Pi_{\mu\nu}^{(b)}(p)=\f{N_{f} g^2}{4a^2}\int_t
tr\lb\lc \f{(s_t^2)^k\sla{s}_t}{w(t)+M(t)} + 1\rc
 \f{1}{2w(t)}\right.\nonumber\\
&&\times\lb X_{2\mu\nu}(t,t,-pa,pa) - \f{X_0(t)}{w(t)}
X^\dagger_{2\mu\nu}(t,t,-pa,pa)\f{X_0(t)}{w(t)}\right.\nonumber\\
&&-\f{1}{(w(t)+w(t+pa))^2}\nonumber\\
&&\times\lc X_{1\mu}(t,pa+t,-pa)
X^\dagger_{1\nu}(pa+t,t,pa)X_0(t)\right.\nonumber\\
&&+X_{1\mu}(t,pa+t,-pa)
X_0^\dagger(pa+t)X_{1\nu}(pa+t,t,pa)\nonumber\\
&&+X_0(t)X^\dagger_{1\mu}(t,pa+t,-pa)
X_{1\nu}(pa+t,t,pa)\nonumber\\
&&-\f{2w(t)+w(pa+t)}{w(t)^2w(pa+t)}\nonumber\\
&&\left.\times X_0(t)X^\dagger_{1\mu}(t,pa+t,-pa)
X_0(pa+t)X^\dagger_{1\nu}(pa+t,t,pa)X_0(t)\rc\nonumber\\
&& +X_{2\nu\mu}(t,t,pa,-pa) - \f{X_0(t)}{w(t)}
X^\dagger_{2\nu\mu}(t,t,pa,-pa)\f{X_0(t)}{w(t)}\nonumber\\
&&-\f{1}{(w(t)+w(t-pa))^2}\nonumber\\
&&\times\lc X_{1\nu}(t,t-pa,pa)
X^\dagger_{1\mu}(t-pa,t,-pa)X_0(t)\right.\nonumber\\
&&+X_{1\nu}(t,t-pa,pa)
X_0^\dagger(t-pa)X_{1\mu}(t-pa,t,-pa)\nonumber\\
&&+X_0(t)X^\dagger_{1\nu}(t,t-pa,pa)
X_{1\mu}(t-pa,t,-pa)\nonumber\\
&&-\f{2w(t)+w(t-pa)}{w(t)^2w(t-pa)}\nonumber\\
&&\left.\left.\left.\times X_0(t)X^\dagger_{1\nu}(t,t-pa,pa)
X_0(t-pa)X^\dagger_{1\mu}(t-pa,t,-pa)X_0(t)\rc\rb\rb,\nonumber\\
\end{eqnarray}  
where $\int_t\equiv \int_{-\pi}^\pi d^4 t/(2\pi)^4$. In the
above two equations $w,\, X_0,\, X_{1\mu},\, X_{2\mu\nu}$ are
appropriately redefined according to the rescaling of $t_\mu$.
For example, 
\begin{equation}
w(t)=\sqrt{\l(s^2_t\r)^{2k+1}
+\lc\l(r\sum_{\rho}(1-\cos t_\rho )\r)^{2k+1}
- \l(m_0\r)^{2k+1}\rc^2}
\end{equation}
where $s_t^2=\sum_{\mu}\sin^2 t_\mu$.

We first want to show that there are no
 nonlocal divergences of the forms $p^2/(a^2 p^2)^{n}$
or  $p_\mu p_\nu/(a^2 p^2)^n \> (n\ge 2)$. For this 
purpose we confirm that Eq.(\ref{pia}) and Eq.(\ref{pib}) are
not singular for $p=0$. Setting $p=0$ in these equations, we have
\begin{eqnarray}
&&\Pi_{\mu\nu}^{(a)}(0)=\f{-N_{f} g^2}{8a^2}\int_t
\f{1}{4w(t)^2}\nonumber\\
&&tr\lb\lc \f{(s_t^2)^k\sla{s}_t}{w(t)+M(t)} + 1
\rc\right.\nonumber\\
&&\times \lc X_{1\mu}(t,t,0)-\f{X_0(t)}{w(t)}
X_{1\mu}^\dagger(t,t,0)\f{X_0(t)}{w(t)}\rc
\nonumber\\
&&\times \lc \f{(s_{t}^2)^k\sla{s}_{t}}{w(t)+M(t)} + 1\rc
\nonumber\\
&&\left.\times \lc X_{1\nu}(t,t,0)-\f{X_0(t)}{w(t)}
X_{1\nu}^\dagger(t,t,0)\f{X_0(t)}{w(t)}\rc\rb.\nonumber\\ 
\end{eqnarray}
and
\begin{eqnarray}
&&\Pi_{\mu\nu}^{(b)}(0)=\f{N_{f} g^2}{4a^2}\int_t
tr\lb\lc \f{(s_t^2)^k\sla{s}_t}{w(t)+M(t)} + 1\rc
 \f{1}{2w(t)}\right.\nonumber\\
&&\times\lb X_{2\mu\nu}(t,t,0,0) - \f{X_0(t)}{w(t)}
X^\dagger_{2\mu\nu}(t,t,0,0)\f{X_0(t)}{w(t)}\right.\nonumber\\
&&-\f{1}{4w(t)^2}\nonumber\\
&&\times\lc X_{1\mu}(t,t,0)
X^\dagger_{1\nu}(t,t,0)X_0(t)\right.\nonumber\\
&&+X_{1\mu}(t,t,0)
X_0^\dagger(t)X_{1\nu}(t,t,0)\nonumber\\
&&+X_0(t)X^\dagger_{1\mu}(t,t,0)
X_{1\nu}(t,t,0)\nonumber\\
&&-\f{3}{w(t)^2}\nonumber\\
&&\left.\times X_0(t)X^\dagger_{1\mu}(t,t,0)
X_0(t)X^\dagger_{1\nu}(t,t,0)X_0(t)\rc\nonumber\\
&& +X_{2\nu\mu}(t,t,0,0) - \f{X_0(t)}{w(t)}
X^\dagger_{2\nu\mu}(t,t,0,0)\f{X_0(t)}{w(t)}\nonumber\\
&&-\f{1}{4w(t)^2}\nonumber\\
&&\times\lc X_{1\nu}(t,t,0)
X^\dagger_{1\mu}(t,t,0)X_0(t)\right.\nonumber\\
&&+X_{1\nu}(t,t,0)
X_0^\dagger(t)X_{1\mu}(t,t,0)\nonumber\\
&&+X_0(t)X^\dagger_{1\nu}(t,t,0)
X_{1\mu}(t,t,0)\nonumber\\
&&-\f{3}{w(t)^2}\nonumber\\
&&\left.\left.\left.\times X_0(t)X^\dagger_{1\nu}(t,t,0)
X_0(t)X^\dagger_{1\mu}(t,t,0)X_0(t)\rc\rb\rb.\nonumber\\
\end{eqnarray}  
Now on the basis of the expressions of $w,\, X_0,\, X_{1\mu}$ and
 $X_{2\mu\nu}$
and the fact that there are no doublers, the possible singularity
 may occur only around the region $t\simeq 0$ in each integral. 
Only the fermion propagators can exhibit singular behavior for
$t\simeq 0$. The leading singularity in 
$t$ in $\Pi_{\mu\nu}^{(a)}(0)$ vanishes as
\begin{eqnarray}
\Pi_{\mu\nu}^{(a)}(0)&\approx&\int_{t^2<\delta^2}tr\lb
\f{\sla{t}}{(t^2)^{k+1}}(t^2)^k
\ga_\mu\f{\sla{t}}{(t^2)^{k+1}}(t^2)^k\ga_\nu\rb
\sim 0\quad (\delta\ll 1),
\end{eqnarray}
and similarly the leading singularity in $\Pi_{\mu\nu}^{(b)}(0)$ 
vanishes as
\begin{eqnarray}
\Pi_{\mu\nu}^{(b)}(0)&\approx&\int_{t^2<\delta^2}tr\lb
\f{\sla{t}}{(t^2)^{k+1}}\ga_\mu\ga_\nu(t^2)^{k-1}\sla{t}\rb
\sim 0\quad (\delta\ll 1).
\end{eqnarray} 
Higher order terms in $t$ are obviously non-singular. Since
both $\Pi_{\mu\nu}^{(a)}(0)$ and $\Pi_{\mu\nu}^{(b)}(0)$ are
not singular, $\Pi_{\mu\nu}(p)$ does not have the non-local 
divergences of the forms $p^2/(a^2 p^2)^{n}$
or $p_\mu p_\nu/(a^2 p^2)^n \> (n\ge 2)$.

There may still exist the quadratic divergence in 
$\Pi_{\mu\nu}(p)$. From Eq.(\ref{Ward id}), the form of
the quadratic divergence for small $p_{\mu}$ is
\begin{eqnarray}
\frac{1}{a^{2}}(\delta_{\mu\nu}
-\frac{\tilde{p}_{\mu}\tilde{p}_{\nu}}{\tilde{p}^{2}})C,
\end{eqnarray}
with a constant $C$.
We have already established that the 
$p_{\mu}\rightarrow 0$ limit of $\Pi_{\mu\nu}(p)$ is 
well-defined, which excludes the singular term 
$\tilde{p}_{\mu}\tilde{p}_{\nu}/\tilde{p}^{2}$; 
this term depends on the direction of 
the approach $p_{\mu}\rightarrow 0$. We thus conclude $C=0$, 
namely, the quadratic divergences cancel between diagrams (a) 
and (b).

Next we confirm that there are no divergences of the 
structure such as $a^2p^2\times \infty$, etc, which 
vanish in the naive continuum limit. These unusual divergences 
, which may be termed as infrared singularities, may occur in 
our treatment of $H_{(2k+1)}$ which corresponds to 
a higher derivative theory on the lattice.
These divergences, if they should exist,  could
appear in the integration region around $t\simeq 0$ and could
remain even for arbitrarily small $p$. Therefore we evaluate
$\Pi^{(a)}_{\mu\nu}(p)$(\ref{pia}) and 
$\Pi^{(b)}_{\mu\nu}(p)$(\ref{pib}) for
$t^2<\delta^2$ and $ap\sim 0$. After a straightforward
calculation, we obtain
\begin{eqnarray}
\Pi^{(a)}_{\mu\nu}(p)&\simeq&-\f{N_f g^2}{2a^2}
\int_{t^2<\delta^2}tr\lb \f{1}{(i\sla{t})^{2k+1}}\left(
\sum_{l+m=2k}(i\sla{t})^l i\gamma_\mu(i(\sla{t}+\sla{p}a))^m
\right)\right.\nonumber\\
&&\left.\times\f{1}{(i(\sla{t}+\sla{p}a))^{2k+1}}
\left(\sum_{l+m=2k}(i(\sla{t}+\sla{p}a))^l i\gamma_\nu
(i\sla{t})^m\right)\rb, \\
\Pi^{(b)}_{\mu\nu}(p)&\simeq&\f{N_f g^2}{2a^2}
\int_{t^2<\delta^2}tr\lb \f{1}{(i\sla{t})^{2k+1}}\lc
\sum_{l+m+n=2k-1}(i\sla{t})^l i\gamma_\mu(i(\sla{t}+\sla{p}a))^m
i\gamma_\nu(i\sla{t})^n \right.\right.\nonumber\\
&&\left.\left. +\sum_{l+m+n=2k-1}(i\sla{t})^l i\gamma_\nu
(i(\sla{t}-\sla{p}a))^m i\gamma_\mu(i\sla{t})^n\rc\rb.  
\end{eqnarray}
These amplitudes (a) and (b) separately could contain infrared
singularities. The cancellation between the amplitudes (a) and 
(b) further takes place as:
\begin{eqnarray}
\Pi^{(a)}_{\mu\nu}(p)+\Pi^{(b)}_{\mu\nu}(p)&\simeq&
-\f{N_f g^2}{2a^2}(2k+1)
\int_{t^2<\delta^2}tr\lb\f{1}{i\sla{t}}i\gamma_\mu
\f{1}{i(\sla{t}+\sla{p}a)}i\gamma_\nu\rb.\nonumber\\
\end{eqnarray}  
This final expression, which has the same structure as that in 
 continuum theory, means that there are no divergences such 
as $a^2p^2\times \infty$, etc.
\\

Finally, we investigate the logarithmic divergence. From the
above analyses, we know that $\Pi_{\mu\nu}(p)$ does not have the 
divergences of the negative power in $a$. Therefore if there is 
the logarithmic
divergence in $\Pi_{\mu\nu}(p)$, it appears from the singular
part in the integral for $a\to 0$ and thus the singular part 
should appear in the integration region around $t\simeq 0$. 
We first evaluate $\Pi_{\mu\nu}^{(a)}(p)$. There are
several ways to extract the logarithmic 
divergence~\cite{Karsten:1981wd}
\cite{Kawai:1981ja}. Here we use
the procedure discussed in the paper by Karsten and 
Smit~\cite{Karsten:1981wd}. First,
the denominators of the propagator are combined using Feynman
parameters and the integration variables are shifted $t_\mu
\to t_\mu -p_\mu ax$ as follows,
\begin{eqnarray}
\label{log1}
&&\Pi_{\mu\nu}^{(a)}(p)=\f{-N_{f} g^2}{8a^2}
\f{\Gamma(4k+2)}{\left(\Gamma(2k+1)\right)^2}
\int_{-\pi +pax}^{\pi+pax}\f{d^4 t}{(2\pi)^4}
\int_0^1 dx\,\f{x^{2k}(1-x)^{2k}}{\lc w(t-pax)+w(t+pa(1-x))\rc^2}
\nonumber\\
&&\times \f{1}{\lb \alpha(1-x)
+ \beta x\rb^{4k+2}}
\nonumber\\
&&\times tr\lb (s_{t-pax}^2)^k\sla{s}_{t-pax}
\lc X_{1\mu}(t-pax,t+pa(1-x),-pa)\right.\right.\nonumber\\
&&\left. -\f{X_0(t-pax)}{w(t-pax)}
X_{1\mu}^\dagger(t-pax,t+pa(1-x),-pa)
\f{X_0(t+pa(1-x))}{w(t+pa(1-x))}\rc
\nonumber\\
&&\times (s_{t+pa(1-x)}^2)^k\sla{s}_{t+pa(1-x)}
\lc X_{1\nu}(t+pa(1-x),t-pax,pa)\right.\nonumber\\
&&\left.\left.-\f{X_0(t+pa(1-x))}{w(t+pa(1-x))}
X_{1\nu}^\dagger(t+pa(1-x),t-pax,pa)
\f{X_0(t-pax)}{w(t-pax)}\rc\rb,\nonumber\\ 
\end{eqnarray}  
where\footnote{We used the Feynman's formula:
\begin{eqnarray}
\f{1}{\alpha^{2k+1}\beta^{2k+1}}=\f{\Gamma(4k+2)}{\Gamma(2k+1)
\Gamma(2k+1)}\int^1_0 dx\>\f{x^{2k}(1-x)^{2k}}{\lb\alpha x +
\beta(1-x)\rb^{4k+2}}.\nonumber
\end{eqnarray}}
\begin{eqnarray}
 \alpha&\equiv& \lc w(t-pax) + M(t-pax)\rc^{1/(2k+1)},\\
 \beta&\equiv&\lc w(t+pa(1-x)) + M(t+pa(1-x))\rc^{1/(2k+1)}.
\end{eqnarray}

Then we split the integration domain into two regions as follows
\begin{eqnarray}
\int_{t^2} = \int_{t^2<\delta^2} + \int_{t^2>\delta^2}\qquad
\delta\ll 1,
\end{eqnarray}
and we evaluate the $t^2<\delta^2$ part in the continuum limit,
ignoring the $t^2>\delta^2$ part which does not contain 
divergence. Eq.(\ref{log1}) is the complicated integral
including sines and cosines. However for $t^2<\delta^2$ and
$a\to 0$ with fixed small $p_{\mu}$ we can expand both the 
denominator and the numerator
of Eq.(\ref{log1}) separately in powers of $t$ and $a$, and we 
have
\begin{eqnarray}
&&\Pi_{\mu\nu}^{(a)}(p)\simeq \f{-N_{f} g^2}{2a^2}
\f{\Gamma(4k+2)}{\left(\Gamma(2k+1)\right)^2}
\int_{t^2<\delta^2}\f{d^4 t}{(2\pi)^4}
\int_0^1 dx\, \f{x^{2k}(1-x)^{2k}}{\lc t^2 + p^2a^2x(1-x)
\rc^{4k+2}}
\nonumber\\
&&\times tr\lb i\lb i(\sla{t}-\sla{p}ax)\rb^{2k+1}
\lc \sum_{0\le l\le 2k}i\lb i(\sla{t}-\sla{p}ax)\rb^{l}i\ga_\mu 
\lb i(\sla{t}+\sla{p}a(1-x))\rb^{2k-l}
\rc\right.\nonumber\\
&&\left. \times i\lb i(\sla{t}+\sla{p}a(1-x))\rb^{2k+1}
\lc \sum_{0\le m\le 2k}i\lb i(\sla{t}+\sla{p}a(1-x))\rb^{m}
i\ga_\nu 
\lb i(\sla{t}-\sla{p}ax)\rb^{2k-m}
\rc\rb.\nonumber\\
\end{eqnarray}
We next evaluate $\Pi_{\mu\nu}^{(b)}(p)$ in the similar way
and we obtain
\begin{eqnarray}
&&\Pi_{\mu\nu}^{(b)}(p)\simeq \f{-N_{f} g^2}{2a^2}
\f{\Gamma(4k+2)}{\left(\Gamma(2k+1)\right)^2}
\int_{t^2<\delta^2}\f{d^4 t}{(2\pi)^4}
\int_0^1 dx\, \f{x^{2k}(1-x)^{2k}}{\lc t^2 + p^2a^2x(1-x)
\rc^{4k+2}}
\nonumber\\
&&\times tr\lb\lb i(\sla{t}+\sla{p}a(1-x))\rb^{4k+2}
i\lb i(\sla{t}-\sla{p}ax)\rb^{2k+1}\right.\nonumber\\
&&\times \lc \sum_{0\le l+m\le 2k-1}i\lb i(\sla{t}-\sla{p}ax)
\rb^{l}i\gamma_\mu\right.\nonumber\\
&&\left.\times \lb i(\sla{t}+\sla{p}a(1-x))\rb^{m}i\gamma_\nu
\lb i(\sla{t}-\sla{p}ax)\rb^{2k-1-l-m}\rc\nonumber\\
&&+ \lb i(\sla{t}-\sla{p}ax)\rb^{4k+2}
i\lb i(\sla{t}+\sla{p}a(1-x))\rb^{2k+1}\nonumber\\
&&\times \lc \sum_{0\le l+m\le 2k-1}
i\lb i(\sla{t}+\sla{p}a(1-x))\rb^{l}
i\gamma_\nu\right.\nonumber\\
&&\left.\left.\times \lb i(\sla{t}-\sla{p}ax)\rb^{m}i\gamma_\mu
\lb i(\sla{t}+\sla{p}a(1-x))\rb^{2k-1-l-m}\rc\rb.\nonumber\\
\end{eqnarray}
By this way
$\Pi_{\mu\nu}(p)\equiv\Pi_{\mu\nu}^{(a)}(p)
+\Pi_{\mu\nu}^{(b)}(p)$ is written as
\begin{eqnarray}
&&\Pi_{\mu\nu}(p)
\simeq \f{-N_{f} g^2}{2a^2}
\f{\Gamma(4k+2)}{\left(\Gamma(2k+1)\right)^2}
\int_{t^2<\delta^2}\f{d^4 t}{(2\pi)^4}
\int_0^1 dx\, \f{x^{2k}(1-x)^{2k}}{\lc t^2 + p^2a^2x(1-x)
\rc^{4k+2}}\nonumber\\
&&\times (2k+1)(t-pax)^{4k}(t+pa(1-x))^{4k}
tr\lb (\sla{t}-\sla{p}ax)\gamma_\mu
(\sla{t}+\sla{p}a(1-x))\gamma_\nu\rb
\end{eqnarray}
The singular part corresponding
to the logarithmic divergence is obtained from the leading part
in $t$ and $a$. Noting the spherical symmetry of the integral
and dropping ${\cal O}(a^3)$ terms in the numerator,
the singular part is given by
\begin{eqnarray}
&&\Pi_{\mu\nu}(p)\simeq \f{-N_{f} g^2}{2a^2}
\int_{t^2<\delta^2}\f{d^4 t}{(2\pi)^4}
\int_0^1 dx\, \f{x^{2k}(1-x)^{2k}}
{\lc t^2 + p^2a^2x(1-x)\rc^{4k+2}}
\nonumber\\
&&\times 4(2k+1)\lb -\f{1}{2}(t^2)^{4k+1}\delta_{\mu\nu}
\right.\nonumber\\
&&+(t^2)^{4k}p^2a^2\delta_{\mu\nu}\lc -(x^2+(1-x)^2)
\left(\f{4}{3}k^2+\f{4}{3}k\right)+x(1-x)
\left(\f{8}{3}k^2+2k+1\right)\rc\nonumber\\
&&\left.+(t^2)^{4k}p_\mu p_\nu a^2\lc (x^2+(1-x)^2)
\left(\f{4}{3}k^2+\f{4}{3}k\right)-x(1-x)
\left(\f{8}{3}k^2+4k+2\right)\rc\rb\nonumber\\
\end{eqnarray}
After some calculations, the term proportional to $\log p^2a^2$
is obtained as (by restoring the factor $\delta^{AB}$) 
\begin{eqnarray}
(2k+1)\delta^{AB}\f{N_{f} g^2}{24\pi^2}
(p^2\delta_{\mu\nu}-p_\mu p_\nu)\log p^2a^2.
\end{eqnarray}
Combined with the general analysis (2.14) in the previous 
section, we conclude that the divergent part of the gauge field 
renormalization factor arising from fermion one-loop diagrams
for the general Dirac operator $D=(\g5/a)H$ is given by 
\begin{eqnarray}
Z_A = 1+\f{N_f g^2}{24\pi^2}\log \mu^2a^2,
\end{eqnarray}
where $\mu$ is the renormalization scale. This factor indeed 
reproduces the correct result for the QCD-type continuum 
theory [23][24]. 

Incidentally, the result (3.32) could also be directly obtained 
from (3.23), which corresponds to $2k+1$ times the vacuum 
polarization tensor generated by a conventional massless 
fermion. 

\section{The vacuum polarization tensor for $H$ with $k=1$}

In this section we calculate the one-loop fermion contribution
to the vacuum polarization tensor $\Pi_{\mu\nu}$ on the basis 
of $H$ with the simplest case  $k=1$. We perform essentially 
the same analysis as in the previous section.

Feynman diagrams for the vacuum polarization with a 
fermion loop are shown in Fig.1, and the 
necessary Feynman rules are given in Appendix B.                     

The amplitude corresponding to Fig.1(a) is given by (for QCD
with $N_{f}$ flavors)
\begin{eqnarray}
&&\Pi_{\mu\nu}^{(a)}(p)=\f{-g^2}{a^2}\frac{N_{f}}{2}\delta^{AB}
\int_t
\f{1}{\alpha(t,t+p)^2}\nonumber\\
&&\times tr\lb D_0^{-1}(t)\lc D(t,t+p)\g5 
H_{(3)1\mu}(t,t+p,-p)\right.\right.\nonumber\\
&&\left. - \g5 H_0(t)
(\g5 H_{(3)1\mu}(t,t+p,-p))^\dagger \g5 H_0(t+p)\rc
\nonumber\\
&&\times D_0^{-1}(t+p)\nonumber\\
&&\times \lc D(t+p,t)\g5 H_{(3)1\nu}(t+p,t,p)
\right.\nonumber\\
&&\left.\left.- \g5 H_0(t+p)
(\g5 H_{(3)1\nu}(t+p,t,p))^\dagger \g5 H_0(t)\rc\rb.
\nonumber\\
\end{eqnarray}

The amplitude corresponding to Fig.1(b) is
\begin{eqnarray}
&&\Pi_{\mu\nu}^{(b)}(p)=
\f{g^2}{a}\frac{N_{f}}{2}\delta^{AB}\int_t tr\lb D_0^{-1}(t)
\f{1}{\alpha(t,t)}\right.\nonumber\\
&&\times \lc D(t,t)\g5 H_{(3)2\mu\nu}(t,t,-p,p)\right.
\nonumber\\
&&-\g5 H_0(t)
(\g5 H_{(3)2\mu\nu}(t,t,-p,p))^\dagger \g5 H_0(t)
\nonumber\\
&&-H_0(t)^2\g5 H_{1\mu}(t,t+p,-p)
(\g5 H_{1\nu}(t+p,t,p))^\dagger \g5 H_0(t)\nonumber\\
&&-2H_0(t)^2\g5 H_{1\mu}(t,t+p,-p)
(\g5 H_0(t+p))^\dagger
\g5 H_{1\nu}(t+p,t,p)\nonumber\\
&&-H_0(t)^2\g5 H_0(t)(\g5 H_{1\mu}(t,t+p,-p))^\dagger
\g5 H_{1\nu}(t+p,t,p)\nonumber\\
&&+\g5 H_0(t)(\g5 H_{1\mu}(t,t+p,-p))^\dagger
\g5 H_0(t+p)(\g5 H_{1\nu}(t+p,t,p))^\dagger \g5 H_0(t)
\nonumber\\
&&\left.\left.+(p,\mu \leftrightarrow -p,\nu)\rc\rb,
\end{eqnarray} 
where
\begin{eqnarray}
\g5 H_{1\nu}(t+p,t,p)&=&\f{1}{\alpha(t,t+p)}\lc
D(t+p,t)\g5 H_{(3)1\nu}(t+p,t,p)\right.\nonumber\\
&&\left.- \g5 H_0(t+p)
(\g5 H_{(3)1\nu}(t+p,t,p))^\dagger \g5 H_0(t)\rc\nonumber\\
\end{eqnarray}

We first show that
the Ward identity for $\Pi_{\mu\nu}$ holds in this case also.
>From the analysis in Subsection 3.1, we obtain
\begin{eqnarray}
&&\sum_{\nu}\tilde{p}_\nu\left(\g5 H_{(3)2\mu\nu}(t,t,-p,p)
+\g5 H_{(3)2\nu\mu}(t,t,p,-p)\right)\nonumber\\
&&=\g5 H_{(3)1\mu}(t,t+p,-p) -\g5 H_{(3)1\mu}(t-p,t,-p).
\nonumber\\
\end{eqnarray}
Using these relations $\sum_{\nu}\tilde{p}_\nu
\Pi_{\mu\nu}^{(b)}(p)$ is written as
\begin{eqnarray}
&&\sum_{\nu}\tilde{p}_\nu
\Pi_{\mu\nu}^{(b)}(p) = \f{g^2}{a}\frac{N_{f}}{2}\delta^{AB}
\int_t
tr\lb D_0^{-1}(t)\f{1}{\alpha(t,t)}\right.\nonumber\\
&&\times \lc D(t,t)(\g5 H_{(3)1\mu}(t,t+p,-p)-
\g5 H_{(3)1\mu}(t-p,t,-p)\right.
\nonumber\\
&&-\g5 H_0(t)
(\g5 H_{(3)1\mu}(t,t+p,-p)-\g5 
H_{(3)1\mu}(t-p,t,-p))^\dagger \g5 H_0(t)
\nonumber\\
&&+H_0(t)^2\g5 H_{1\mu}H_0(t+p)H_0(t)
-I\times\g5 H_{1\mu}\nonumber\\
&&-2H_0(t)^2\g5 H_0(t)H_{1\mu}^\dagger
H_0(t+p)+D(t,t+p)\g5 H_0(t)H_{1\mu}H_0(t)\nonumber\\
&&-H_0(t)^2\g5 H_{1\mu}^\prime 
H_0(t-p)H_0(t)
+I^\prime\times\g5 
H_{1\mu}^\prime \nonumber\\
&&\left.\left.+2H_0(t)^2\g5 H_0(t)
H_{1\mu}^{\dagger\prime}
H_0(t-p)-D(t,t-p)\g5 H_0(t)
H_{1\mu}^\prime H_0(t)\rc\rb\nonumber\\
\end{eqnarray}
where $'\prime'$ means $p\rightarrow -p$ and 
\begin{eqnarray}
I&=&-3H_0(t)^{4}+2H_0(t)^{2}D(t,t+p).
\end{eqnarray}
Noting Eq.(\ref{2}), $\sum_{\nu}\tilde{p}_\nu
\Pi_{\mu\nu}^{(b)}(p)$ is rewritten as follows,
\begin{eqnarray}
\label{wardpib}
&&\sum_{\nu}\tilde{p}_\nu
\Pi_{\mu\nu}^{(b)}(p) = \f{g^2}{a}\frac{N_{f}}{2}\delta^{AB}
\int_t
tr\lb D_0^{-1}(t)\f{1}{\alpha(t,t)}\g5 \right.\nonumber\\
&&\times \lc 2H_0(t)^{2}(H_{(3)1\mu}(t,t+p,-p)-
H_{(3)1\mu}(t-p,t,-p)\right.
\nonumber\\
&&-H_0(t)
(H_{(3)1\mu}(t,t+p,-p)- 
H_{(3)1\mu}(t-p,t,-p))H_0(t)
\nonumber\\
&&+3H_0(t)^{4}H_{1\mu}(t,t+p,-p)
-2H_0(t)^{2}H_{(3)1\mu}(t,t+p,-p)\nonumber\\
&& +H_0(t)H_{(3)1\mu}(t,t+p,-p)
H_0(t)\nonumber\\
&&-3H_0(t)^{4}H_{1\mu}(t-p,t,-p)
+2H_0(t)^{2}H_{(3)1\mu}(t-p,t,-p)\nonumber\\
&&\left.\left. -H_0(t)H_{(3)1\mu}(t-p,t,-p)
H_0(t)\rc\rb\nonumber\\
&&=\f{g^2}{a}\frac{N_{f}}{2}\delta^{AB}\int_t
tr\lb D_0^{-1}(t)\g5(H_{1\mu}(t,t+p,-p)-H_{1\mu}(t-p,t,-p))\rb
\nonumber\\
\end{eqnarray}
We next calculate $\sum_{\nu}\tilde{p}_\nu
\Pi_{\mu\nu}^{(a)}(p)$,
\begin{eqnarray}
\label{wardpia}
&&\sum_{\nu}\tilde{p}_\nu
\Pi_{\mu\nu}^{(a)}(p)=\f{-g^2}{a^2}\frac{N_{f}}{2}\delta^{AB}
\int_t
\nonumber\\
&&\times tr\lb D_0^{-1}(t)\g5 H_{1\mu}(t,t+p,-p) D_0^{-1}(t+p)
\g5 (H_0(t+p)-H_0(t))\rb\nonumber\\
&&=\f{-g^2}{a}\frac{N_{f}}{2}\delta^{AB}\int_t tr\lb (D_0^{-1}(t)- 
D_0^{-1}(t+p))
\g5 H_{1\mu}(t,t+p,-p) \rb\nonumber\\
&&=\f{-g^2}{a}\frac{N_{f}}{2}\delta^{AB}\int_t tr\lb D_0^{-1}(t)
\g5 (H_{1\mu}(t,t+p,-p)-H_{1\mu}(t-p,t,-p)) \rb.\nonumber\\
\end{eqnarray}
>From Eq.(\ref{wardpia}) and Eq.(\ref{wardpib}), 
one can see that the Ward identity for the vacuum polarization 
tensor  holds.

We next examine the structure of various divergences. 
Rescaling the integration momenta
$t_\mu\to t_\mu/a$ in each amplitude, we obtain (by omitting the 
factor $\delta^{AB}$ from now on) 
\begin{eqnarray}
\label{rescaleHa}
&&\Pi_{\mu\nu}^{(a)}(p)=\f{-N_fg^2}{2a^2}\int_t
\f{1}{\alpha(t,t+pa)^2}\nonumber\\
&&\times tr\lb D_0^{-1}(t)\lc D(t,t+pa)\g5 
H_{(3)1\mu}(t,t+pa,-pa)\right.\right.\nonumber\\
&&\left. - \g5 H_0(t)
(\g5 H_{(3)1\mu}(t,t+pa,-pa))^\dagger \g5 H_0(t+pa)\rc
\nonumber\\
&&\times D_0^{-1}(t+pa)\nonumber\\
&&\times \lc D(t+pa,t)\g5 H_{(3)1\nu}(t+pa,t,pa)
\right.\nonumber\\
&&\left.\left.- \g5 H_0(t+pa)
(\g5 H_{(3)1\nu}(t+pa,t,pa))^\dagger \g5 H_0(t)\rc\rb.
\nonumber\\
\end{eqnarray}
and
\begin{eqnarray}
\label{resclaleHb}
&&\Pi_{\mu\nu}^{(b)}(p)=
\f{N_fg^2}{2a^2} \int_t tr\lb D_0^{-1}(t)
\f{1}{\alpha(t,t)}\right.\nonumber\\
&&\times \lc D(t,t)\g5 H_{(3)2\mu\nu}(t,t,-pa,pa)\right.
\nonumber\\
&&-\g5 H_0(t)
(\g5 H_{(3)2\mu\nu}(t,t,-pa,pa))^\dagger \g5 H_0(t)
\nonumber\\
&&-H_0(t)^2\g5 H_{1\mu}(t,t+pa,-pa)
(\g5 H_{1\nu}(t+pa,t,pa))^\dagger \g5 H_0(t)\nonumber\\
&&-2H_0(t)^2\g5 H_{1\mu}(t,t+pa,-pa)
(\g5 H_0(t+pa))^\dagger
\g5 H_{1\nu}(t+pa,t,pa)\nonumber\\
&&-H_0(t)^2\g5 H_0(t)(\g5 H_{1\mu}(t,t+pa,-pa))^\dagger
\g5 H_{1\nu}(t+pa,t,pa)\nonumber\\
&&+\g5 H_0(t)(\g5 H_{1\mu}(t,t+pa,-pa))^\dagger
\g5 H_0(t+pa)(\g5 H_{1\nu}(t+pa,t,pa))^\dagger \g5 H_0(t)
\nonumber\\
&&\left.\left.+(p,\mu \leftrightarrow -p,\nu)\rc\rb.
\end{eqnarray} 
First, we want to show that $\Pi_{\mu\nu}^{(a)}(p)$ and
$\Pi_{\mu\nu}^{(b)}(p)$ are finite and well defined for $p=0$,
 and thus the divergent terms of the forms $p^2/(a^2p^2)^{n+1}
\delta_{\mu\nu}$ and $p_\mu p_\nu/(a^2p^2)^{n+1}$ with $n\geq 0$ 
do not appear.
Setting $p=0$ in these expressions, we have
\begin{eqnarray}
&&\Pi_{\mu\nu}^{(a)}(0)=\f{-N_fg^2}{2a^2}\int_t
\f{1}{\alpha(t,t)^2}\nonumber\\
&&\times tr\lb D_0^{-1}(t)\lc D(t,t)\g5 
H_{(3)1\mu}(t,t,0)\right.\right.\nonumber\\
&&\left. - \g5 H_0(t)
(\g5 H_{(3)1\mu}(t,t,0))^\dagger \g5 H_0(t)\rc
\nonumber\\
&&\times D_0^{-1}(t)\nonumber\\
&&\times \lc D(t,t)\g5 H_{(3)1\nu}(t,t,0)
\right.\nonumber\\
&&\left.\left.- \g5 H_0(t)
(\g5 H_{(3)1\nu}(t,t,0))^\dagger \g5 H_0(t)\rc\rb.
\nonumber\\
\end{eqnarray}
and
\begin{eqnarray}
&&\Pi_{\mu\nu}^{(b)}(0)=
\f{N_fg^2}{2a^2}\int_t tr\lb D_0^{-1}(t)
\f{1}{\alpha(t,t)}\right.\nonumber\\
&&\times \lc D(t,t)\g5 H_{(3)2\mu\nu}(t,t,0,0)\right.
\nonumber\\
&&-\g5 H_0(t)
(\g5 H_{(3)2\mu\nu}(t,t,0,0))^\dagger \g5 H_0(t)
\nonumber\\
&&-H_0(t)^2\g5 H_{1\mu}(t,t,0)
(\g5 H_{1\nu}(t,t,0))^\dagger \g5 H_0(t)\nonumber\\
&&-2H_0(t)^2\g5 H_{1\mu}(t,t,0)
(\g5 H_0(t))^\dagger
\g5 H_{1\nu}(t,t,0)\nonumber\\
&&-H_0(t)^2\g5 H_0(t)(\g5 H_{1\mu}(t,t,0))^\dagger
\g5 H_{1\nu}(t,t,0)\nonumber\\
&&+\g5 H_0(t)(\g5 H_{1\mu}(t,t,0))^\dagger
\g5 H_0(t)(\g5 H_{1\nu}(t,t,0))^\dagger \g5 H_0(t)
\nonumber\\
&&\left.\left.+(\mu \leftrightarrow \nu)\rc\rb.
\end{eqnarray} 
The singularity may occur around the region $t\simeq 0$ 
in each integral. However, the leading order part in $t$ in
$\Pi_{\mu\nu}^{(a)}(0)$ and $\Pi_{\mu\nu}^{(b)}(0)$ vanish as
\begin{eqnarray}
\Pi_{\mu\nu}^{(a)}(0)&\simeq& -\f{N_fg^2}{2a^2}
\int_{t^2<\delta^2}\f{1}{(t^2)^{4}}\times
tr\lb \f{\sla{t}}{t^2}(t^2)^{2}\gamma_\mu
\f{\sla{t}}{t^2}(t^2)^{2}\gamma_\nu\rb \simeq 0,\nonumber\\
\Pi_{\mu\nu}^{(b)}(0)&\simeq& \f{N_fg^2}{2a^2}
\int_{t^2<\delta^2}tr \lb \f{\sla{t}}{t^2}\f{1}{(t^2)^{2}}
(t^2)\sla{t}\gamma_\mu\gamma_\nu\rb \simeq 0,
\nonumber
\end{eqnarray}
for $\delta\ll 1$. Since both $\Pi_{\mu\nu}^{(a)}(0)$ 
and $\Pi_{\mu\nu}^{(b)}(0)$ are non-singular, 
$\Pi_{\mu\nu}$  does not contain the non-local divergences.
>From the fact that the $p_\mu\to 0$ limit of  
$\Pi_{\mu\nu}(p)$ is well-defined and finite and that 
the Ward identity holds, 
we also conclude that the possible quadratic divergences cancel
between diagrams (a) and (b). See also the analysis in Section 3.

Next we confirm that there are no divergences of the structure
such as $a^2p^2\times \infty$, etc, which vanish in the naive
continuum limit even if they existed. 
For this purpose we evaluate $\Pi_{\mu\nu}^{(a)}(p)$ 
and $\Pi_{\mu\nu}^{(b)}(p)$ for $t^2<\delta^2$ and $ap\sim 0$.
We  thus examine the behaviour of various functions appearing
in these amplitudes for $t^2<\delta^2$ and $ap\sim 0$. 
They are given as follows,
\begin{eqnarray}
&&D_0^{-1}(t)\simeq (2M_0)^{1/3}\f{\sla{t}}{t^2},
\quad \alpha(t,t)\simeq 3\left(\f{1}{2M_0}\right)^{4/3}
(t^2)^{2},\nonumber\\
&&\gamma_5H_0(t)\simeq -\left(\f{1}{2M_0}\right)^{1/3}\sla{t},
\quad H_0(t)^2\simeq \left(\f{1}{2M_0}\right)^{2/3}t^2,
\nonumber\\
&&\gamma_5 H_{1\mu}(t,t+p,-p)\simeq 
-\left(\f{1}{2M_0}\right)^{1/3}\gamma_\mu, \nonumber\\
&&D(t,t+pa)\simeq\left(\f{1}{2M_0}\right)^{2/3}\lc
t^2 + (t+pa)^2 \rc,\nonumber\\
&&\gamma_5H_{(3)2\mu\nu}(t,t,-pa,pa)
\simeq \f{1}{2M_0}\sum_{l+m+n=1}\lc
i(i\sla{t})^l i\gamma_\mu (i(\sla{t}+\sla{p}a))^m i\gamma_\nu
(i\sla{t})^n\rc,\nonumber
\end{eqnarray}
where all the higher order terms are non-singular.
Using these expressions, $\Pi_{\mu\nu}^{(a)}(p)$ 
and $\Pi_{\mu\nu}^{(b)}(p)$ for $t^2<\delta^2$ and $ap\sim 0$ 
are expressed as 
\begin{eqnarray}
\Pi_{\mu\nu}^{(a)}(p)&\simeq&-\f{N_fg^2}{2a^2}\int_{t^2<\delta^2}
\f{1}{t^2(t+pa)^2}tr\lb \sla{t}\gamma_\mu(\sla{t}+\sla{p}a)
\gamma_\nu\rb,\nonumber\\
\Pi_{\mu\nu}^{(b)}(p)&\simeq&\f{N_fg^2}{2a^2}\int_{t^2<\delta^2}
tr\lb \f{\sla{t}}{t^2}\f{1}{t^2}f(t,pa)\rb,\nonumber
\end{eqnarray}
where $f(t,pa)$ is a non-singular function of $t$ and $pa$.
Since $\Pi_{\mu\nu}^{(b)}(p)$ vanishes in this limit and 
the expression of $\Pi_{\mu\nu}^{(a)}(p)$ has the
same structure as that in continuum theory, we conclude that
there are no (unusual) divergences such as 
$a^2p^2\times\infty$, etc.,in the vacuum polarization tensor. 

Finally, we investigate the logarithmic divergence.
>From the above analyses, the logarithmic divergence
in $\Pi_{\mu\nu}(p)$ appears from the singular part
in the integral for $a\to 0$ and $t\simeq 0$.
Since $\Pi_{\mu\nu}^{(b)}(p)$ is non-singular in this limit,
we consider only the amplitude $\Pi_{\mu\nu}^{(a)}(p)$. 
In Eq.(\ref{rescaleHa}), we use the  Feynman's parameter and
shift the integration variables $t_\mu\to t_\mu-ap_\mu x$. We 
then  evaluate the contribution from the integration region
$t^2<\delta^2$, $\delta\ll 1$ in the continuum limit.
We then have
\begin{eqnarray}
\Pi_{\mu\nu}^{(a)}(p)&\simeq&-\f{N_fg^2}{2a^2}\int_{t^2<\delta^2}
\f{d^4 t}{(2\pi)^4}\int_0^1 dx\>
\f{x^{2}(1-x)^{2}}{\lc t^2+p^2a^2x(1-x)\rc^{6}}
\f{\Gamma(6)}{\Gamma(3)^2}\nonumber\\
&&\times((t-pax)^2)^2((t+pa(1-x))^2)^2
tr\lb (\sla{t}-\sla{p}ax)\gamma_\mu(\sla{t}+\sla{p}a(1-x))
\gamma_\nu\rb.\nonumber\\
\end{eqnarray}
After some calculations, the term proportional to 
$\log a^2p^2$ is extracted as follows,
\begin{eqnarray}
\f{N_fg^2}{24\pi^2}(p^2\delta_{\mu\nu}-p_\mu p_\nu)\log a^2p^2.
\end{eqnarray} 
This expression agrees with the one expected from the general 
analysis in Section 2.

\section{Discussion}

We have studied a perturbative aspect of a general class of 
Dirac operators. To avoid the excessive complications, we 
examined the simplest diagrams of the one-loop fermion 
correction to the gauge field self-energy tensor. This quantity 
is related to the one-loop
$\beta$ function and also to the Weyl anomaly. We have confirmed
that the perturbative analysis gives the correct result for 
any $k\geq 1$ by using the relation (2.14), in 
accord with the general analysis of  Weyl anomaly. This correct 
result is consistent with our previous analyses of the locality
 of the general operator $H_{(2k+1)}$ and the locality domain of
 $|F_{\mu\nu}|$ for $H_{(2k+1)}$ [12]: Also, our result does not 
contradict the general perturbative analysis of lattice theory 
in [26] if one remembers the locality properties of $H_{(2k+1)}$.
We have also confirmed the relation (2.14) for the simplest case
$k=1$ by evaluating the self-energy correction in terms of the 
operator $H$ itself. 

When combined with the analysis of chiral anomaly[8], our 
present analysis shows that all the local anomalies are 
properly reproduced by our general class of operators $D$. 
These analyses give some confidence in the treatment of 
the fermionic determinant
\begin{equation}
\det H=(\det H_{(2k+1)})^{1/(2k+1)}
\end{equation}
in the possible application to QCD, for example.

At the same time, we recognized that infrared divergences 
may generally appear in the intermediate stages of perturbative 
calculations for finite $a$, 
which should cancel in the final result. This treatment of 
infrared divergences in perturbation theory is quite 
 tedious in our generalized operator $D$. To avoid the 
infrared complications, the (nonperturbative) Wilsonian effective
 action, which is supposed to be free of infrared complications, 
is expected to be essential for the general operator $D$. 
As for the perturbative treatment of  fully dynamical gauge 
field such as in the 
one-loop correction to the fermion self-energy, some auxiliary 
regulator such as the dimensional regulator may become necessary
[22] for a reliable treatment of infrared divergences.

\appendix

\section{Feynman rules for the general $H_{(2k+1)}$}

 We derive the Feynman rules for $H_{(2k+1)}$ 
theory (not $H$ itself) to calculate the vacuum polarization at 
one-loop level on the basis of (2.14). $H_{(2k+1)}$ has been 
defined by (1.6), (1.7) and (1.8).
We expand $H_{(2k+1)}$ up to the second order in the coupling
constant $g$ as follows,
\begin{eqnarray}
\label{1}
H_{(2k+1)}&=&H_{(2k+1)0} + gH_{(2k+1)1} + g^2 H_{(2k+1)2} 
+{\cal O}(g^{3}).
\end{eqnarray} 
For this purpose we first need to expand $D_W^{(2k+1)}(x,y)$ in 
 $g$. $D_W^{(2k+1)}(x,y)$ up to the second order in $g$ is given 
by
\be
D_W^{(2k+1)}(x,y)&=&a^4\intp\intq\, e^{ipx-iqy}\lb X_0(p)
\delta_P(p-q) + X_1(p,q) + X_2(p,q)\right.\nonumber\\
&&\left. + O(g^{3})\rb,
\ee
where
\begin{eqnarray}
X_0(p)&=&i\left(\f{i}{a}\ga_\mu\sin p_\mu a\right)^{2k+1} + 
\left(\f{r}{a}\sum_\mu (1-\cos p_\mu a)\right)^{2k+1} 
- \left(\f{m_0}{a}\right)^{2k+1}\nonumber\\
&=&-\l(\f{\sin^2 p_\rho a}{a^2}\r)^k
\f{\gamma_\rho\sin p_\rho a}{a} + M(p),\\
X_1(p,q)&\equiv&\sum_{\mu}\inttt\delta_P(p-t-q)gA_\mu(t)
X_{1\mu}(p,q,t),\\
X_2(p,q)&\equiv&\sum_{\mu,\nu}\int_{-\f{\pi}{a}}^{\f{\pi}{a}}
\f{d^4k_1}{(2\pi)^4}
\int_{-\f{\pi}{a}}^{\f{\pi}{a}}\f{d^4k_2}{(2\pi)^4}\delta_P
(p-q-k_1
-k_2)g^2A_\mu(k_1)A_\nu(k_2)\nonumber\\
&&\times X_{2\mu\nu}(p,q,k_1,k_2).\nonumber\\
\end{eqnarray}
Here we defined 
\begin{eqnarray}
&&X_{1\mu}(p,q,t)=\sum_{l+m=2k}\lc i
\left(\f{i}{a}\ga_\rho\sin p_\rho a\right)^l\left(i\ga_\mu\cos(q
+\f{t}{2})_\mu a\right)
\left(\f{i}{a}\ga_\rho\sin q_\rho a\right)^{m} 
\right.\nonumber\\
&&\left.+\left(\f{r}{a}\sum_\rho(1-\cos p_\rho a)\right)^{l}
\left(r\sin(q+\f{t}{2}
)_\mu a\right)\left(\f{r}{a}\sum_\rho (1-\cos q_\rho a)
\right)^{m}\rc,
\end{eqnarray}
and 
\begin{eqnarray}
&&X_{2\mu\nu}(p,q,k_1,k_2)=
\sum_{l+m+n=2k-1}\lc 
i\left(\f{i}{a}\ga_\rho\sin p_\rho a\right)^l
\left(i\ga_\mu\cos(p
-\f{k_1}{2})_\mu a\right)\right.\nonumber\\
&&\times\left(\f{i}{a}\ga_\rho\sin (p-k_1)_\rho a
\right)^m\left(i\ga_\nu\cos(q
+\f{k_2}{2})_\nu a\right)\left(\f{i}{a}\ga_\rho\sin q_\rho a
\right)^n
\nonumber\\
&&+
\left(\f{r}{a}\sum_\rho(1-\cos p_\rho a)\right)^{l}
\left(r\sin(p-\f{k_1}{2}
)_\mu a\right)\nonumber\\
&&\left.\times\left(\f{r}{a}\sum_\rho(1-\cos(p-k_1)_\rho a)
\right)^m
\left(r\sin(q+\f{k_2}{2}
)_\nu a\right)\left(\f{r}{a}\sum_\rho(1-\cos q_\rho a)\right)^{n}
\rc\nonumber\\
&&+\f{1}{2}\sum_{l+m=2k}\lc i
\left(\f{i}{a}\ga_\rho\sin p_\rho a\right)^l\left(-ia\ga_\mu
\delta_{\mu\nu}
\sin(q+\f{k_1+k_2}{2})_\mu a\right)
\left(\f{i}{a}\ga_\rho\sin q_\rho a\right)^{m}\right.\nonumber\\
&& \left.+
\left(\f{r}{a}\sum_\rho(1-\cos p_\rho a)\right)^{l}
\left(ar\delta_{\mu\nu}\cos(q+\f{k_1+k_2}{2}
)_\mu a\right)
\left(\f{r}{a}\sum_\rho (1-\cos q_\rho a)
\right)^{m}\rc\nonumber\\
\end{eqnarray} 
where
\begin{eqnarray}
&&M(p)= \l(\f{r}{a}\sum_{\rho}(1-\cos p_\rho a)\r)^{2k+1}
- \l(\f{m_0}{a}\r)^{2k+1},
\end{eqnarray}
and $l,m,n$ are the nonnegative integers 
and $\delta_P$ is the 
periodic lattice delta function. $A_\mu(k)$ is defined  by
\be
A_\mu(x)=\intk\,A_\mu(k)e^{ik(x-\f{a\hatmu}{2})},
\ee
and has the properties $A_\mu^\dagger(k) = A_\mu(-k),\quad
A_\mu(k+\f{2\pi}{a}l)=(-1)^{l_\mu}A_\mu(k)\quad(l:integer)$.  

Next we want to expand the factor 
$1/\sqrt{(D_W^{(2k+1)})^\dagger
(D_W^{(2k+1)})}$ in $H_{(2k+1)}$ (\ref{H_{(2k+1)}}). But 
it is very complicated to perform the weak coupling expansion 
of $1/\sqrt{(D_W^{(2k+1)})^\dagger
(D_W^{(2k+1)})}$ directly, if not impossible. Therefore we use 
the following identity:
\be
\f{1}{\sqrt{X^\dagger X}} = \int^\infty_{-\infty}\f{dt}{\pi}\f{1}
{t^2 +  X^\dagger X}.
\ee
The weak coupling expansion of the 
integrand on the right-hand side can be readily performed. 
After some calculations, we obtain
\begin{eqnarray}
\label{wkH2k+1}
H_{(2k+1)}&=&a^4\intp\intq e^{ipx-iqy}\f{1}{2}\g5
\lb\l(1+\f{X_0(p)}{\wp} \r)\delta_P(p-q)\right.\nonumber\\
&& + \lc \f{1}{\omega(p) + \omega(q)}\rc 
\lc X_1(p,q) - \f{X_0(p)}{\omega(p)}X^\dagger_1(p,q)
\f{X_0(q)}{\omega(q)}\rc
\nonumber\\
&&+\lc \f{1}{\omega(p) + \omega(q)}\rc 
\lc X_2(p,q) - \f{X_0(p)}{\omega(p)}X^\dagger_2(p,q)
\f{X_0(q)}{\omega(q)}\rc
\nonumber\\
&& + \int_{t}
\lc \f{1}{\omega(p) + \omega(q)}\rc
\lc \f{1}{\omega(p) + \omega(t)}\rc
\lc \f{1}{\omega(t) + \omega(q)}\rc\nonumber\\
&&\times\lc- X_0(p)X^\dagger_1(p,t)X_1(t,q)\right.\nonumber\\
&& \quad -X_1(p,t)X_0^\dagger(t)X_1(t,q) - X_1(p,t)X^\dagger_1
(t,q)X_0(q)\nonumber\\
&&\left.\left. \quad +\f{\omega(p) + \omega(q) +\omega(t)}
{\omega(p)
\omega(t)\omega(q)}X_0(p)X^\dagger_1(p,t)X_0(t)X^\dagger_1(t,q)
X_0(q)\rc
 +{\cal O}(g^{3})\rb,\nonumber\\
\end{eqnarray}
where
\begin{eqnarray}
\wp&=&\sqrt{\l(\f{1}{a^2}\sin^2 p_\rho a\r)^{2k+1}
+\lc\l(\f{r}{a}\sum_{\rho}(1-\cos p_\rho a)\r)^{2k+1}
- \l(\f{m_0}{a}\r)^{2k+1}\rc^2},\nonumber\\
\end{eqnarray}
and $X^\dagger \equiv \g5 X \g5$.
We write
$H_{(2k+1)0}(x,y),H_{(2k+1)1}(x,y)$ and $H_{(2k+1)2}(x,y)$ 
as follows,
\begin{eqnarray}
&&H_{(2k+1)0}(x,y)=a^4\int_p e^{ipx-ipy} H_{(2k+1)0}(p),\\
&&H_{(2k+1)1}(x,y)=a^4\sum_{\mu}\intpqt e^{ipx-iqy}
\delta_P(p-q-t)
A_\mu(t)H_{(2k+1)1\mu}(p,q,t),\\
&&H_{(2k+1)2}(x,y)=a^4\sum_{\mu\nu}\int_{p,q,k_1,k_2} e^{ipx-iqy}
\delta_P(p-q-k_1-k_2)
A_\mu(k_1)A_\nu(k_2)H_{(2k+1)2\mu\nu}(p,q,k_1,k_2),\nonumber\\
\end{eqnarray}
where $\int_p \equiv \intpp$. Then from Eq.(\ref{wkH2k+1}), 
$H_{(2k+1)0}(p)$, 
$H_{(2k+1)1\mu}(p,q,t)$ and\\ $H_{(2k+1)2\mu\nu}(p,q,k_1,k_2)$ 
can be written as
\begin{eqnarray}
&&\label{prop}H_{(2k+1)0}(p) = \f{1}{2}\g5\l
(1+\f{X_0(p)}{\wp}\r),\\
&&\label{ver1}H_{(2k+1)1\mu}(p,q,t)=\f{1}{2}\g5\f{1}{\wp+\wq}
\nonumber\\
&&\times \lc X_{1\mu}(p,q,t) - \f{X_0(p)}{\wp}X^\dagger_{1\mu}
(p,q,t)
\f{X_0(q)}{\wq}\rc,\\
&&\label{ver2}
H_{(2k+1)2\mu\nu}(p,q,k_1,k_2)=\f{1}{2}\g5\f{1}{\wp+\wq}
\nonumber\\
&&\times\lb  X_{2\mu\nu}(p,q,k_1,k_2) - \f{X_0(p)}{\wp}
X^\dagger_{2\mu\nu}(p,q,k_1,k_2)\f{X_0(q)}{\wq}\right.\nonumber\\
&&-\f{1}{(\wp+\wpk)(\wpk+\wq)}\nonumber\\
&&\times\lc X_{1\mu}(p,p-k_1,k_1)
X^\dagger_{1\nu}(q+k_2,q,k_2)X_0(q)\right.\nonumber\\
&&+X_{1\mu}(p,p-k_1,k_1)
X_0^\dagger(p-k_1)X_{1\nu}(q+k_2,q,k_2)\nonumber\\
&&+X_0(p)X^\dagger_{1\mu}(p,p-k_1,k_1)
X_{1\nu}(q+k_2,q,k_2)\nonumber\\
&&-\f{\wp+\wpk+\wq}{\wp\wpk\wq}\nonumber\\
&&\left.\left.\times X_0(p)X^\dagger_{1\mu}(p,p-k_1,k_1)
X_0(p-k_1)
X^\dagger_{1\nu}(q+k_2,q,k_2)X_0(q)\rc\rb.\nonumber\\
\end{eqnarray} 
Note that $q=p-t$ in $H_{(2k+1)1\mu}(p,q,t)$ and 
$q=p-k_1-k_2$ in $H_{(2k+1)2\mu\nu}(p,q,k_1,k_2)$.

>From this weak coupling expansion we can derive the Feynman
rules for $H_{(2k+1)}$ theory, which are 
necessary for the one-loop 
analyses. To make the structure of the divergences in
one-loop amplitudes explicit
we derive the Feynman rules for 
$\f{1}{a^{2k+1}}H_{(2k+1)} = (\g5 D)^{2k+1}$.
Using Eq.(\ref{prop}), the fermion propagator $D_0^{-1}(p)$,
where $D_0^{-1}(p)$ stands for the inverse of free  
$(\g5 D)^{2k+1}$, is written as
\begin{eqnarray}
D_0^{-1}(p)=\left(\f{\left(s_p^2\right)^k\sla{s}_p}{w(p)+M(p)}
+a^{2k+1}
\right)\g5
\end{eqnarray}
where $\sla{s}_p\equiv\sum_{\mu}\ga_\mu\sin ap_\mu$ and
$s_p^2\equiv\sum_{\mu}\sin^2 ap_\mu$. 
Using Eq.(\ref{ver1}), we assign the following factor to the 
fermion-gauge field three-point vertex depicted in Fig.2  
\begin{eqnarray}
-\f{g}{2a^{2k+1}}T^A_{ba}\g5\f{1}{\wp+\wq}
\lc X_{1\mu}(p,q,t) - \f{X_0(p)}{\wp}X^\dagger_{1\mu}(p,q,t)
\f{X_0(q)}{\wq}\rc
\end{eqnarray}
where $T^A$ are $SU(N)$ generators. Using Eq.(\ref{ver2}),
we assign the following factor to the
fermion-gauge field four-point vertex depicted in Fig.3 
\begin{eqnarray}
&&-\f{g^2}{2a^{2k+1}}(T^AT^B)_{ba}\g5\f{1}{\wp+\wq}
\nonumber\\
&&\times\lb  X_{2\mu\nu}(p,q,k_1,k_2) - \f{X_0(p)}{\wp}
X^\dagger_{2\mu\nu}(p,q,k_1,k_2)\f{X_0(q)}{\wq}\right.\nonumber\\
&&-\f{1}{(\wp+\wpk)(\wpk+\wq)}\nonumber\\
&&\times\lc X_{1\mu}(p,p-k_1,k_1)
X^\dagger_{1\nu}(q+k_2,q,k_2)X_0(q)\right.\nonumber\\
&&+X_{1\mu}(p,p-k_1,k_1)
X_0^\dagger(p-k_1)X_{1\nu}(q+k_2,q,k_2)\nonumber\\
&&+X_0(p)X^\dagger_{1\mu}(p,p-k_1,k_1)
X_{1\nu}(q+k_2,q,k_2)\nonumber\\
&&-\f{\wp+\wpk+\wq}{\wp\wpk\wq}\nonumber\\
&&\left.\left.\times X_0(p)X^\dagger_{1\mu}(p,p-k_1,k_1)
X_0(p-k_1)X^\dagger_{1\nu}(q+k_2,q,k_2)X_0(q)\rc\rb\nonumber\\
&&+(A,\mu,k_1 \leftrightarrow B,\nu,k_2)
\end{eqnarray}
where we have imposed the Bose symmetry for
gauge fields.

%The gauge field action is chosen as Wilson's plaquette 
%action 
%\begin{equation}
%  S_G = \frac{1}{g^2}\sum_m\sum_{\mu\nu} 
%{\rm Tr} \left(1- U_{\mu\nu}(m) \right),
%\end{equation}
%where $U_{\mu\nu}(m)=U_\mu(m)U_\nu(m+\hat \mu) 
%U_\mu(m+\hat \nu)^\dagger U_\nu(m)^\dagger $.
%The gluon propagator is given by 
%\begin{eqnarray}
%G_{\mu\nu}(k) &=& \f{1}{\tilde k^2}\left( \delta_{\mu\nu} 
%- (1 - \alpha)
%\f{\tilde k_\mu \tilde k_\nu}{\tilde k^2}\right),
%\end{eqnarray}
%where $\tilde k_\mu =(2/a)\sin ak_\mu/2$, and $\alpha$ is the
%gauge fixing parameter .

\begin{figure}[htbp]
\begin{center}
%WinTpicVersion2.15
\unitlength 0.1in
\begin{picture}(22.20,13.79)(13.70,-25.90)
% VECTOR 2 0 3 0
% 2 1410 2790 1990 2790
% 
\special{pn 8}%
\special{pa 1410 2390}%
\special{pa 1990 2390}%
\special{fp}%
\special{sh 1}%
\special{pa 1990 2390}%
\special{pa 1923 2370}%
\special{pa 1937 2390}%
\special{pa 1923 2410}%
\special{pa 1990 2390}%
\special{fp}%
% VECTOR 2 0 3 0
% 2 1980 2790 2980 2790
% 
\special{pn 8}%
\special{pa 1980 2390}%
\special{pa 2980 2390}%
\special{fp}%
\special{sh 1}%
\special{pa 2980 2390}%
\special{pa 2913 2370}%
\special{pa 2927 2390}%
\special{pa 2913 2410}%
\special{pa 2980 2390}%
\special{fp}%
% LINE 2 0 3 0
% 2 2980 2790 3590 2790
% 
\special{pn 8}%
\special{pa 2980 2390}%
\special{pa 3590 2390}%
\special{fp}%
% STR 2 0 3 0
% 3 1610 2890 1610 2990 1 0
% p,a
\put(16.1000,-25.9000){\makebox(0,0)[lt]{p,a}}%
% STR 2 0 3 0
% 3 3200 2890 3200 2990 1 0
% q,b
\put(32.0000,-25.9000){\makebox(0,0)[lt]{q,b}}%
% LINE 2 2 3 0
% 2 2480 2785 2480 1611
% 
\special{pn 8}%
\special{pa 2480 2385}%
\special{pa 2480 1211}%
\special{dt 0.045}%
\special{pa 2480 1211}%
\special{pa 2480 1212}%
\special{dt 0.045}%
% VECTOR 2 0 3 0
% 2 2340 2120 2340 1866
% 
\special{pn 8}%
\special{pa 2340 1720}%
\special{pa 2340 1466}%
\special{fp}%
\special{sh 1}%
\special{pa 2340 1466}%
\special{pa 2320 1533}%
\special{pa 2340 1519}%
\special{pa 2360 1533}%
\special{pa 2340 1466}%
\special{fp}%
% STR 2 0 3 0
% 3 1370 1810 1370 1910 1 0
% $A,\mu,t$
\put(13.7000,-15.1000){\makebox(0,0)[lt]{$A,\mu,t$}}%
\end{picture}%
\end{center}
\label{fig1}
\caption{Fermion-gauge field three-point vertex}
\end{figure}
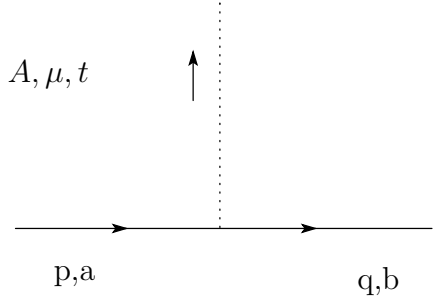
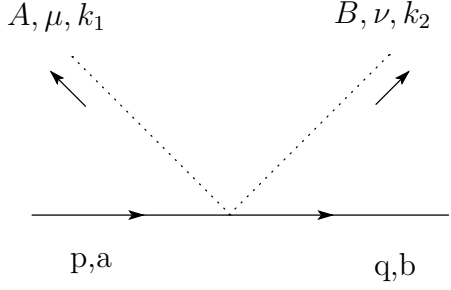
\begin{figure}[htbp]
\begin{center}
%WinTpicVersion2.15
\unitlength 0.1in
\begin{picture}(23.10,13.20)(12.80,-25.90)
% VECTOR 2 0 3 0
% 2 1410 2790 1990 2790
% 
\special{pn 8}%
\special{pa 1410 2390}%
\special{pa 1990 2390}%
\special{fp}%
\special{sh 1}%
\special{pa 1990 2390}%
\special{pa 1923 2370}%
\special{pa 1937 2390}%
\special{pa 1923 2410}%
\special{pa 1990 2390}%
\special{fp}%
% VECTOR 2 0 3 0
% 2 1980 2790 2980 2790
% 
\special{pn 8}%
\special{pa 1980 2390}%
\special{pa 2980 2390}%
\special{fp}%
\special{sh 1}%
\special{pa 2980 2390}%
\special{pa 2913 2370}%
\special{pa 2927 2390}%
\special{pa 2913 2410}%
\special{pa 2980 2390}%
\special{fp}%
% LINE 2 0 3 0
% 2 2980 2790 3590 2790
% 
\special{pn 8}%
\special{pa 2980 2390}%
\special{pa 3590 2390}%
\special{fp}%
% STR 2 0 3 0
% 3 1610 2890 1610 2990 1 0
% p,a
\put(16.1000,-25.9000){\makebox(0,0)[lt]{p,a}}%
% STR 2 0 3 0
% 3 3200 2890 3200 2990 1 0
% q,b
\put(32.0000,-25.9000){\makebox(0,0)[lt]{q,b}}%
% LINE 2 2 3 0
% 2 2450 2790 1620 1960
% 
\special{pn 8}%
\special{pa 2450 2390}%
\special{pa 1620 1560}%
\special{dt 0.045}%
\special{pa 1620 1560}%
\special{pa 1620 1560}%
\special{dt 0.045}%
% LINE 2 2 3 0
% 2 2440 2790 3280 1950
% 
\special{pn 8}%
\special{pa 2440 2390}%
\special{pa 3280 1550}%
\special{dt 0.045}%
\special{pa 3280 1550}%
\special{pa 3280 1550}%
\special{dt 0.045}%
% VECTOR 2 0 3 0
% 2 3210 2210 3380 2040
% 
\special{pn 8}%
\special{pa 3210 1810}%
\special{pa 3380 1640}%
\special{fp}%
\special{sh 1}%
\special{pa 3380 1640}%
\special{pa 3319 1673}%
\special{pa 3342 1678}%
\special{pa 3347 1701}%
\special{pa 3380 1640}%
\special{fp}%
% VECTOR 2 0 3 0
% 2 1690 2220 1510 2040
% 
\special{pn 8}%
\special{pa 1690 1820}%
\special{pa 1510 1640}%
\special{fp}%
\special{sh 1}%
\special{pa 1510 1640}%
\special{pa 1543 1701}%
\special{pa 1548 1678}%
\special{pa 1571 1673}%
\special{pa 1510 1640}%
\special{fp}%
% STR 2 0 3 0
% 3 1280 1580 1280 1680 1 0
% $A,\mu,k_1$
\put(12.8000,-12.8000){\makebox(0,0)[lt]{$A,\mu,k_1$}}%
% STR 2 0 3 0
% 3 2990 1570 2990 1670 1 0
% $B,\nu,k_2$
\put(29.9000,-12.7000){\makebox(0,0)[lt]{$B,\nu,k_2$}}%
\end{picture}%
\end{center}
\label{fig2}
\caption{Fermion-gauge field four-point vertex}
\end{figure}

In Section 3 of the present paper, we calculate the vacuum 
polarization tensor  at one-loop level by using these Feynman 
rules.

\section{Feynman rules for the operator $H$ with k=1}

 In this Appendix we derive the Feynman rules for the operator 
$H$ to calculate the vacuum polarization 
at one-loop level. For simplicity we consider the case with 
$k=1$. 
For the Feynman rules for the operator $H_{(3)}$,
 we refer to Appendix A. 
We expand $H$ and $H_{(3)}$ up to the second order in the 
coupling constant $g$ as follows,
\begin{eqnarray}
H&=&H_0 + gH_1 +g^2 H_2 + {\cal O}(g^3)\\
D&=&\f{1}{a}\g5 H_0 + \f{1}{a}\g5 g H_1 + 
\f{1}{a}\g5 g^2 H_2 + {\cal O}(g^3)\nonumber\\
\nonumber\\
H_{(3)}&=&H_{(3)0} + gH_{(3)1} + g^2 H_{(3)2} 
+{\cal O}(g^3)\nonumber\\
\label{1}
&=&\left(H_0 + gH_1 +g^2 H_2 + {\cal O}(g^3)\right)^{3}\quad 
(=H^3)\nonumber\\
&=&H_0^{3}+g\sum_{0\le m\le 2}H_0^m H_1 H_0^{2-m}\nonumber\\
&&+g^2\left(\sum_{0\le m \le 2}H_0^m H_2 H_0^{2-m}
+\sum_{0\le l+m \le 1}H_0^l H_1 H_0^m H_1 H_0^{1-l-m}\right)
\nonumber\\
&&+{\cal O}(g^3)
\end{eqnarray} 
Now we want to know the expressions of $H_0,H_1,H_2$. 
We have obtained $H_0$ in our previous paper[12][9]. In 
momentum space $H_0$ is written as
\begin{eqnarray}
\label{4}H_0(p)&=&\g5 \l(\f{1}{2}\r)^{\f{2}{3}}
\l(\f{1}{\wp}\r)^{\f{2}{3}}\lc(\wp + M(p))^{\f{2}{3}}
-(\wp - M(p))^{\f{1}{3}}\f{\sla{s_p}}{a}\rc,\nonumber\\
&=&\g5 \l(\f{1}{2\wp}\r)^{\f{2}{3}}
\l(\f{1}{\wp + M(p)}\r)^{\f{1}{3}}\lc -\l(\f{s_p^2}{a^2}
\r)\f{\sla{s_p}}{a} + \wp + M(p)\rc,\nonumber\\
\end{eqnarray}
where $\sla{s_p}\equiv \sum_{\mu}\gamma_\mu \sin ap_\mu$ and
$s_p^2\equiv \sum_{\mu}\sin^2 ap_\mu$.
One can easily check that $H_{(3)0}(p) = (H_0(p))^{3}$.
 
Next we derive the expression of $H_1$. We write $H_1$ as 
follows,
\begin{eqnarray}
H_1&=&a^4\sum_{\mu}\intpqt e^{ipx-iqy}\delta_P(p-q-t)
A_\mu(t)H_{1\mu}(p,q,t),
\end{eqnarray} 
where $\intpqt \equiv \intp\intq\intt$. Using Eq.(\ref{1}) 
at first order in $g$, we obtain
\begin{eqnarray}
\label{2}H_{(3)1\mu}(p,q,t)&=&\sum_{0\le m\le 2}H_0(p)^m 
H_{1\mu}(p,q,t)H_0(q)^{2-m}\nonumber\\
&=&D(p,q)H_{1\mu}(p,q,t) + H_0(p)H_{1\mu}(p,q,t)
H_0(q),\nonumber\\
\end{eqnarray}
where
\begin{eqnarray}
D(p,q)&=&H_0(p)^{2}
+H_0(q)^{2},\\
H_0(p)^2 &=& \lc\f{1}{2}\l(1+\f{M(p)}{\wp}\r)\rc^{\f{1}{3}}.
\end{eqnarray} 
 Now we consider an ansatz as
\begin{eqnarray}
\label{5}H_{1\mu}(p,q,t)&=&\f{1}{\alpha}
\l(D(p,q) H_{(3)1\mu}(p,q,t) 
- H_0(p)H_{(3)1\mu}(p,q,t)H_0(q)\r).\nonumber\\
\end{eqnarray}
Substituting this ansatz for $H_{1\mu}$ into Eq.(\ref{2}),
we easily obtain the expression of $\alpha$ as
\begin{eqnarray}
\label{6}\alpha(p,q)&=& 
D^2(p,q) - H_0^2(p)H_0^2(q)\nonumber\\
&=&\sum_{0\le m\le 2}H_0(p)^{2m} H_0(q)^{2(2-m)}.
\end{eqnarray}
Thus we have obtained the expression of $H_{1}$.

Further we derive the expression of $H_2$, performing 
the similar procedure as deriving $H_1$. 
We first write $H_{2}$ as follows,
\begin{eqnarray}
H_2&=&a^4\sum_{\mu\nu}\int_{p,q,k_1,k_2} e^{ipx-iqy}
\delta_P(p-q-k_1-k_2)
A_\mu(k_1)A_\nu(k_2)H_{2\mu\nu}(p,q,k_1,k_2).\nonumber\\
\end{eqnarray}
Using Eq.(\ref{1}) at the second order in $g$, we obtain
\begin{eqnarray}
\label{3}
&&H_{(3)2\mu\nu}(p,q,k_1,k_2)=\sum_{0\le m\le 2}
H_0(p)^m H_{2\mu\nu}(p,q,k_1,k_2)H_0(q)^{2-m}\nonumber\\
&&+\sum_{0\le l+m\le 1}H_0(p)^l H_{1\mu}(p,p-k_1,k_1)
H_0(p-k_1)^m H_{1\nu}(p-k_1,q,k_2) H_0(q)^{1-l-m}\nonumber\\
&&=D(p,q)H_{2\mu\nu}(p,q,k_1,k_2)\nonumber\\
&&+H_0(p)H_{2\mu\nu}(p,q,k_1,k_2)H_0(q)\nonumber\\
&&+H_{1\mu}(p,p-k_1,k_1)H_{1\nu}(p-k_1,q,k_2)H_0(q)
\nonumber\\
&&+H_{1\mu}(p,p-k_1,k_1)H_0(p-k_1)H_{1\nu}(p-k_1,q,k_2)
\nonumber\\
&&+H_0(p)H_{1\mu}(p,p-k_1,k_1)H_{1\nu}(p-k_1,q,k_2)
\end{eqnarray}
Now we define 
\begin{eqnarray}
H^\prime_{(3)2\mu\nu}(p,q,k_1,k_2)&\equiv&
H_{(3)2\mu\nu}\nonumber\\
&&-\lc H_{1\mu}H_{1\nu}H_0+ H_{1\mu}H_0 H_{1\nu}+ 
H_0 H_{1\mu}H_{1\nu}\rc.\nonumber\\
\end{eqnarray} 
Using $H^\prime_{(3)2\mu\nu}$, Eq.(\ref{3}) is written as
\begin{eqnarray}
H_{(3)2\mu\nu}^\prime&=&DH_{2\mu\nu} +H_0(p)H_{2\mu\nu}H_0(q).
\end{eqnarray}
The structure of this equation is the same as that of 
Eq.(\ref{2}). Therefore $H_{2\mu\nu}$ is written as
\begin{eqnarray}
H_{2\mu\nu}&=& \f{1}{\alpha(p,q)}\l(D(p,q)H_{(3)2\mu\nu}^\prime
- H_0(p)H_{(3)2\mu\nu}^\prime H_0(q) \r).\nonumber\\
\end{eqnarray}  
Finally we obtain the expression of $H_{2\mu\nu}$ in momentum 
space as follows,
\begin{eqnarray}
\label{7}
H_{2\mu\nu}(p,q,k_1,k_2)&=&\f{1}{\alpha(p,q)}\lc D(p,q)
H_{(3)2\mu\nu}(p,q,k_1,k_2)\right.\nonumber\\
&&-H_0(p)H_{(3)2\mu\nu}(p,q,k_1,k_2)H_0(q)
\nonumber\\
&&-H_0(q)^2H_{1\mu}(p,p-k_1,k_1) H_{1\nu}(p-k_1,q,k_2) H_0(q)
\nonumber\\
&&-D(p,q)H_{1\mu}(p,p-k_1,k_1) H_0(p-k_1) H_{1\nu}(p-k_1,q,k_2)
\nonumber\\
&&-H_0(p)^2H_0(p)H_{1\mu}(p,p-k_1,k_1) H_{1\nu}(p-k_1,q,k_2)
\nonumber\\
&&\left.+H_0(p) H_{1\mu}(p,p-k_1,k_1) H_0(p-k_1)
H_{1\nu}(p-k_1,q,k_2) H_0(q)\rc.\nonumber\\
\end{eqnarray}
>From this weak coupling expansion we can derive 
the Feynman rules
for $D$ in the case of
$k=1$, which are 
necessary for the one-loop analysis. Using Eq.(\ref{4}),
the fermion propagator $D_0^{-1}(p)$ is written as
\begin{eqnarray}
D_0^{-1}(p) = \f{aH_0(p)^{2}\l(\f{s_p^2}{a^2}\r) 
\f{\sla{s_p}}{a}}{\wp+M(p)} + aH_0(p)^{2}.
\end{eqnarray}
Using Eq.(\ref{5}) and Eq.(\ref{6}), we assign the following
expression to the 
fermion-gauge field three-point vertex depicted in Fig.2, 
\begin{eqnarray}
&&-\f{g}{a}T^A_{ba}\f{1}{\alpha(p,q)}\nonumber\\
&&\times\lc D(p,q) \g5 H_{(3)1\mu}(p,q,t) 
- \g5 H_0(p)\l(\g5 H_{(3)1\mu}
(p,q,t)\r)^\dagger
\g5 H_0(q)\rc,\nonumber\\
\end{eqnarray}
where $T^A$ are $SU(N)$ generators. Using Eq.(\ref{7}),
we assign the following expression to the fermion-gauge 
four-point vertex depicted in Fig.3,
\begin{eqnarray}
&&-\f{g^2}{a}(T^AT^B)_{ba}\f{1}{\alpha(p,q)}\lc D(p,q)
\g5 H_{(3)2\mu\nu}(p,q,k_1,k_2)\right.\nonumber\\
&&-\g5 H_0(p)
\l(\g5 H_{(3)2\mu\nu}(p,q,k_1,k_2)\r)^\dagger
\g5 H_0(q)
\nonumber\\
&&-H_0(q)^2\nonumber\\
&&\times\g5 H_{1\mu}(p,p-k_1,k_1)
\l(\g5 H_{1\nu}(p-k_1,q,k_2)\r)^\dagger
\g5 H_0(q)
\nonumber\\
&&-D(p,q)\nonumber\\
&&\times\g5 H_{1\mu}(p,p-k_1,k_1)
\l(\g5 H_0(p-k_1)\r)^\dagger
\g5 H_{1\nu}(p-k_1,q,k_2)
\nonumber\\
&&-H_0(p)^2\nonumber\\
&&\times\g5 H_0(p)\l(\g5 H_{1\mu}
(p,p-k_1,k_1)\r)^\dagger
\g5 H_{1\nu}(p-k_1,q,k_2)
\nonumber\\
&&+\g5 H_0(p)\l(\g5 H_{1\mu}
(p,p-k_1,k_1)\r)^\dagger
\g5 H_0(p-k_1)\nonumber\\
&&\left.\times\l(\g5 H_{1\nu}(p-k_1,q,k_2)\r)^\dagger 
\g5 H_0(q)\rc +(A,\mu,k_1\leftrightarrow B,\nu,k_2).
\end{eqnarray}
We perform a one-loop calculation in Section 4 on the basis of 
these Feynman rules.

\end{document}